\documentclass[a4paper,11pt]{article}
\pdfoutput=1

\usepackage{jheppub}

\usepackage{amsmath,amssymb,graphicx}
\usepackage{caption}
\usepackage{subcaption}
\usepackage{multirow}
\usepackage{dcolumn}
\usepackage{color,url}
\usepackage{colordvi}
\usepackage{bm}
\usepackage{siunitx}
\usepackage{slashed}
\usepackage{natbib}
\usepackage{hyperref}
\usepackage{makecell}
\usepackage{pbox}
\usepackage{array}
\newcommand{\bea}{\begin{eqnarray}}
\newcommand{\eea}{\end{eqnarray}}
\newcommand{\beq}{\begin{eqnarray}}
\newcommand{\eeq}{\end{eqnarray}}

\newcommand{\ud}{\mathrm{d}}

\newcommand{\mgamc}{Madgraph5\_aMC@NLO}

\title{\boldmath Constraining rare B decays by $\mu^+\mu^-\to tc$ at future lepton colliders}

\author[a]{Sichun Sun}
\author[b,c]{Qi-Shu Yan}
\author[d]{Xiaoran Zhao}
\author[c,e]{Zhijie Zhao}

\affiliation[a]{School of Physics, Beijing Institute of Technology, Beijing, 100081, China}
\affiliation[b]{School of Physics Sciences, University of Chinese Academy of Sciences, Beijing 100039, China}
\affiliation[c]{Center for Future High Energy Physics, Institute of High Energy Physics, Chinese Academy of Sciences, Beijing 100039, China}
\affiliation[d]{Dipartimento di Matematica e Fisica, Universit{\`a} di Roma Tre and \\
INFN, sezione di Roma Tre, I-00146 Rome, Italy}
\affiliation[e]{Deutsches Elektronen-Synchrotron DESY, Notkestr. 85, 22607 Hamburg, Germany}

\emailAdd{sichunssun@gmail.com}
\emailAdd{yanqishu@ucas.ac.cn}
\emailAdd{xiaoran.zhao@uniroma3.it}
\emailAdd{zhijie.zhao@desy.de}

\preprint{}

\abstract{
Motivated by the recent rare B decays measurements, 
we study the matching procedure of operators $O_9, O_{10}$ in the low energy effective Hamiltonian and operators in the Standard Model effective theory (SMEFT). 
It is noticed that there are more related operators in the SMEFT whose coefficients can not be determined only from the low-energy data from B physics. 
We demonstrate how to determine these coefficients with some new physics models, 
like $Z^\prime$ model and leptoquark models, 
and then consider how to probe these operators of SMEFT at high energy by using the process $\mu^+\mu^-\to tc$ at future muon colliders, 
which can provide complementary information except for  $\mu^+ \mu^- \to b s$ on the underlying models which lead to rare B decay processes. 
We perform a Monte Carlo study (a hadron level analysis) to show how to separate the signal events from the SM background events and estimate the sensitivity to the Wilson coefficients for different models.
}

\begin{document}
\maketitle
\flushbottom

\section{Introduction}
Searching for new physics is the prime target of both the high energy frontier and high precision frontier.
In the rare decay of B mesons, 
long-standing discrepancies were reported between the Standard Model predictions and experimental measurements, 
with a hint of non-lepton flavor universality(LFU), 
especially in the muon-related final states. 
These hints are observed in a $B \rightarrow K \mu^+ \mu^-$, 
$B_s \rightarrow \phi  \mu ^+ \mu^-$, $B_s \rightarrow  \mu ^+ \mu^-$ 
and angular distribution of $B \rightarrow K^* \mu ^+ \mu^-$\cite{LHCb:2021awg,LHCb:2021vsc,LHCb:2017rmj,ATLAS:2018cur,CMS:2019bbr}. 
For the LFU violation, the hints were reported by LHCb\cite{LHCb:2021trn,LHCb:2019hip,LHCb:2017avl,LHCb:2015svh,LHCb:2020lmf,LHCb:2020gog} in the ratio 

\begin{eqnarray}
R_K = \frac{BR(B \rightarrow K \mu^+ \mu^-)}{BR(B \rightarrow K e^+ e^-)}, R_{K^*} = \frac{BR(B \rightarrow K^* \mu^+ \mu^-)}{BR(B \rightarrow K^* e^+ e^-)}
\end{eqnarray}

Although large hadronic uncertainties can enter in some of these absolute branching fractions and angular observables for the Standard Model predictions, 
$R_K, R_{K^*}, B_s \rightarrow  \mu ^+ \mu^- $ are considered relatively theoretically clean. 
The deviation in those measurements might lead to indirect evidence for new physics
~\cite{Allanach:2015gkd,Hiller:2014ula, Altmannshofer:2017fio, Altmannshofer:2014rta,Geng:2017svp,Ciuchini:2019usw,Datta:2019zca,Aebischer:2019mlg,Ciuchini:2020gvn,Jager:2017gal}. 
This picture has suddenly changed with the very recent experimental updates from CMS collaboration on BR($B_{(d,s)} \rightarrow \mu^+\mu^-$) with the full Run 2 data~\cite{CMS-PAS-BPH-21-006}, 
and LHCb analysis of $R_K$ and $R_{K*}$ with the full Run 1 and 2 dataset~\cite{LHCb:2022qnv,LHCb:2022zom}. 
The newly reported measurements are in agreement with the Standard Model values, however, the new physics effects can still come into play due to both theoretical and experimental uncertainties~\cite{Ciuchini:2022wbq,Buras:2023ldz,Greljo:2022jac,Das:2022mmh,Desai:2023jxh,Varzielas:2023qlb,Becirevic:2023aov}. 
While that measurement provides hints into new physics, the exact mechanics(models) behind those hints are still unknown,
and low-energy measurement cannot fully reveal the nature behind that.

On the other hand, by scattering high-energy particles,
collider experiments provide unique opportunities to access underlying UV theories.
Among various current and future colliders\cite{Shiltsev:2019rfl},
a multi-TeV muon collider\cite{Aime:2022flm,MuonCollider:2022xlm} is ideal for such studies.
Being fundamental particles, the entire energy of incoming muons is available to produce short-distance scattering rather than being spread among partons of hadrons,
and thus a 14 TeV muon collider can be as effective as a 100 TeV proton-proton collider\cite{Delahaye:2019omf}.
Such high energy reach strongly benefits new heavy particles searches, such as minimal dark matter models\cite{Han:2020uak,Bottaro:2021snn} searches,
as well as indirect measurement at high energies\cite{Buttazzo:2020uzc}.
Moreover, vector boson fusion processes are found to be important at muon colliders\cite{Costantini:2020stv},
and enable access to difficult parameters such as the Higgs quartic self-coupling\cite{Chiesa:2020awd}.
More importantly, muon colliders have a special feature: the initial states are muons, directly related to those low-energy physics. 
The muon $g-2$ anomaly can be probed directly at muon colliders\cite{Buttazzo:2020ibd}.
In the context of muon $g-2$ anomaly, studies have been performed on testing it under the SMEFT formalism\cite{Buttazzo:2020ibd}, 
and model-exhaustive analysis\cite{Buttazzo:2020ibd,Capdevilla:2020qel,Capdevilla:2021rwo,Capdevilla:2021kcf,Yin:2020afe}.

The multi-TeV reach and better precision advantages of muon colliders, make it a perfect place to probe muon-related B physics in low energy. 
There are already proposals studying $\mu^+\mu^-\to bs$ (here $bs$ denotes both $b\bar{s}$ and $\bar{b}s$) at multi-TeV scale to discuss the impact of low energy rare B decays processes \cite{Altmannshofer:2022xri,Huang:2021nkl,Huang:2021biu}. 
At the current stage, some rare B decay processes can be nicely parameterized by effective four-fermion operators at the B physics scale ($\mu= 4.8$ GeV):
 \begin{eqnarray}
   O_9 &=& (\bar{s}\gamma_\mu P_L b)(\bar{\ell}\gamma^{\mu}\ell),  \\
   O_{10} &=& (\bar{s}\gamma_\mu P_L b)(\bar{\ell}\gamma^{\mu}\gamma_5\ell), \\
   O^{\prime}_9 &=& (\bar{s}\gamma_\mu P_R b)(\bar{\ell}\gamma^{\mu}\ell),\\
   O^{\prime}_{10} &=& (\bar{s}\gamma_\mu P_R b)(\bar{\ell}\gamma^{\mu}\gamma_5\ell),
\end{eqnarray}
For the new physics effects described by these operators, when we go above the weak scale ($\mu=M_Z$ for instance), 
the Wilson coefficients of such two operators depend on the specific models (e.g. leptoquark, scalars,$Z'$,etc). 

In this work, we adopt the same assumption that the new physics scale is around $\mu=35$ TeV and it is challenging to discover the new physics signature at the LHC. 
We also assume that these new physics above the weak scale ($\mu=M_Z$) can be described by the framework of the standard model effective field theory (SMEFT). 
Under appropriate assumptions, it is noteworthy that there is at least one more operator needed in order to match the SMEFT to low energy operators $O_9$ and $O_{10}$. 
These three operators in the SMEFT are given in Eqs. (2.12-2.14). 
Different new physics models can lead to different matching conditions across the weak scale.

Different from the proposal presented in \cite{Altmannshofer:2022xri}, 
where the polarization of muon beams and charge tagging of jets in the final state is assumed, 
in this work, in order to reveal the nature of new physics related to the rare B decays, 
we propose to measure the process $\mu^+\mu^-\to tc$ (same as $bs$,  $tc$ denotes both $t \bar{c}$ and $\bar{t} c$). 
Same final state has been studied for $e^+e^-$ collider in Ref.~\cite{Bar-Shalom:1997ezk,Bar-Shalom:1997htk,Bar-Shalom:1999dtk}.
And the $tc\ell\ell$ interaction has been studied through the $t\to c\ell\ell$ decay in~\cite{Bause:2020auq}. 
We extend these works to muon collider, with a detailed Monte-Carlo simulation at hadron level.

To probe the $tc\ell\ell$ interaction at LHC or future hadron colliders,    
a possible way is considering the decay process $t\to c\ell\ell$.
From the latest version of Review of Particle Physics~\cite{ParticleDataGroup:2022pth}, 
there is no direct search to this process. 
Indirect constraints can come from the measurements of $t\to Zq(q=u,c)$ decay, 
where $Z$ further decays to leptons. 
Currently, the branching ratio of $t\to Zq$ is established at $\mathcal{O}(10^{-4})$~\cite{ATLAS:2015vhj,CMS:2017wcz,ATLAS:2018zsq}.
Probing the $tc\ell\ell$ interaction is a huge challenge at hadron colliders. 

This study shows that the measurement of $tc$ final state can provide complementary information to the process $\mu^+\mu^-\to b s$.
The four fermion operators for $\mu^+\mu^-\to tc$ naturally arise from the operator matching conditions from the low energy effective field theory to the SMEFT, 
since left-handed top and charm quarks form electroweak SU(2) doublets respectively as the partners of the left-handed bottom and strange quarks in SMEFT operators. 
The process $\mu^+\mu^-\to tc$ can also help to distinguish different new physics models, which yield different operator matching conditions. 
We will consider one $Z^\prime$ model and three leptoquark models.
 
The leptoquark particles are predicted in the grand unification models and they can either be scalar or vector bosons. 
Usually, these particles are superheavy (e.g. $10^{13}$ GeV), as required by the experimental data of proton decays. 
Very light leptoquarks (e.g. 1 TeV or a few 10 TeV) are consistent with experimental data if their couplings to the first generation of matter fields are weak. 
Light leptoquarks are also predicted in the Pati-Salam model, where lepton numbers are treated as the fourth color quantum number. 
These light leptoquarks can be accessible even at the LHC and future collider projects. 
Recently, leptoquarks have attracted much attention in order to interpret the previously claimed B anomalies. 
A comprehensive review on the phenomenology of leptoquarks can be found in \cite{Dorsner:2016wpm}.

Our new findings in this work include 
1) The dominant SM background events for the process $\mu^+\mu^-\to tc$ are different from the background of the process $\mu^+\mu^-\to bs$. 
Due to the highly boosted top quark in the final states, jet substructure analysis is crucial to distinguish signal and background events. 
2) In the case that there is no new resonance found in the TeV muon colliders, measurement of the process $\mu^+\mu^-\to tc$, 
can provide crucial information on the potential nature of the new physics which leads to the low energy rare B decay processes data. 
3) It is noticed that the final state $W^\pm jj $ can have a very large cross section (e,g. 100fb with collision energy $\sqrt{s}=10$ TeV), 
and such a final state is mainly from the weak final state radiation processes. 
Suppressing the weak final state radiation might be important for signal findings.

This paper is organized as given below. 
In section~\ref{Sec:EFT}, we demonstrate the relations between the Wilsonian coefficients of the low-energy effective Hamiltonian and those of the SMEFT. 
In section~\ref{Sec:NP}, we present the values of Wilson coefficients of the SMEFT derived from different new physics models. 
These new physics models can accommodate the rare B decay processes data. 
In section~\ref{Sec:result}, we perform a Monte Carlo simulation to explore the sensitivity of future muon colliders to these new physics scenarios. 
We end this work with a few discussions and conclusions. 
In Appendix, we present the renormalization group equations of Wilsonian coefficients in the SMEFT and effective Hamiltonian, respectively. 

\section{Matching and running of different operator bases}
\label{Sec:EFT}

Interestingly, these rare B decay processes can be simultaneously explained in a model-independent way with the effective four-fermion operators. 
In many B-physics studies, new physics effects strongly prefer an effective Hamiltonian 
with Wilson coefficients of dimension 6 interactions at the renormalization scale $\mu =4.8$ GeV ,
\begin{eqnarray}
  \mathcal{H}_{eff} &=& \mathcal{H}^{SM}_{eff}-\mathcal{N}\sum_{\ell=e,\mu}\sum_{i=9,10}\left(c_iO^{bs\ell\ell}_i+c^\prime_i O^{\prime bs\ell\ell}_i\right) + h.c., 
\end{eqnarray} 
where the normalization factor $\mathcal{N}$ is 
\begin{eqnarray}
  \mathcal{N} &=& \frac{4G_F}{\sqrt{2}}V_{tb}V^*_{ts}\frac{e^2}{16\pi^2}. 
\end{eqnarray}
The operators $O_9$ and $O_{10}$ are 
\begin{eqnarray}
   O^{bs\ell\ell}_9 &=& (\bar{s}\gamma_\mu P_L b)(\bar{\ell}\gamma^{\mu}\ell),  \\
   O^{bs\ell\ell}_{10} &=& (\bar{s}\gamma_\mu P_L b)(\bar{\ell}\gamma^{\mu}\gamma_5\ell),
\end{eqnarray}
where $P_L=(1-\gamma_5)/2$ is the left-handed projection operator. 
For $O^\prime_i$, $P_L$ is replaced by right-handed projection operator $P_R=(1+\gamma_5)/2$.

For the purpose of this paper, to study the related operator $O_9$ and $O_{10}$ defined in the low energy B physics scale in higher colliding energy scale, 
we need to treat the running and matching of the operators carefully. 
Especially the subtleties that arise when the energy runs across the weak scale. 
Our study finds that in the energy scale above the weak scale, 
the impact of $O_9$ and $O_{10}$ operators needs to be reparametrized by three different operators, 
rather than two, in the so-called Warsaw basis, known as the Standard model effective theory (SMEFT). 
Here we introduce different bases and coefficient matching as below. 

At the scale below the weak scale, potential new physics effects are described by a low energy effective field theory (LEFT).  
The LEFT Lagrangian with dimension 6 operators can be written as 
\begin{eqnarray}
  \mathcal{L}_{LEFT} &=& \mathcal{L}_{QCD+QED}+ \frac{1}{v^2}\sum_{i}L_iQ_i,  \nonumber 
\end{eqnarray} 
where $L_i$ are the Wilson Coefficients of LEFT, and $v=246$ GeV is the vacuum expectation value. 

In this paper, we use the convention of Ref.~\cite{Jenkins:2017jig}. 
The most relevant operators are 
\begin{eqnarray}
  Q^{V,LL}_{ed}(p,r,s,t) &=& (\bar{e}_{Lp}\gamma^\mu e_{Lr})(\bar{d}_{Ls}\gamma_\mu d_{Lt}),  \label{QVLLed} \\
  Q^{V,LR}_{de}(p,r,s,t) &=& (\bar{d}_{Lp}\gamma^\mu d_{Lr})(\bar{e}_{Rs}\gamma_\mu e_{Rt}),  \label{QVLRde} \\
  Q^{V,LR}_{ed}(p,r,s,t) &=& (\bar{e}_{Lp}\gamma^\mu e_{Lr})(\bar{d}_{Rs}\gamma_\mu d_{Rt}),  \label{QVLRed} \\
  Q^{V,RR}_{ed}(p,r,s,t) &=& (\bar{e}_{Rp}\gamma^\mu e_{Rr})(\bar{d}_{Rs}\gamma_\mu d_{Rt}),  \label{QVRRed}
\end{eqnarray}
where $p,r,s,t$ are generation indices of quark or lepton. 
Since the EW symmetry is broken, here $e$ and $d$ are the lepton field and down-type quark field. 
$L$ and $R$ are the chiral indices of fermions. 

One can derive the relations between $Q_i$ and $O_i$ easily:
\begin{eqnarray}
  Q^{V,LL}_{ed} &=& \frac{1}{2}\left(O_9-O_{10}\right),  \\
  Q^{V,LR}_{de} &=& \frac{1}{2}\left(O_9+O_{10}\right),  \\
  Q^{V,LR}_{ed} &=& \frac{1}{2}\left(O^{\prime}_9-O^{\prime}_{10}\right),  \\
  Q^{V,RR}_{de} &=& \frac{1}{2}\left(O^{\prime}_9+O^{\prime}_{10}\right).   
\end{eqnarray}
So we have 
\begin{eqnarray}
  L^{V,LL}_{ed} &=& \frac{\mathcal{N}v^2}{2}(c_9-c_{10}),  \\
  L^{V,LR}_{de} &=& \frac{\mathcal{N}v^2}{2}(c_9+c_{10}),  \\
  L^{V,LR}_{ed} &=& \frac{\mathcal{N}v^2}{2}(c^{\prime}_9-c^{\prime}_{10}),  \\
  L^{V,RR}_{de} &=& \frac{\mathcal{N}v^2}{2}(c^{\prime}_9+c^{\prime}_{10}). 
\end{eqnarray}
The constraints of $c_9$ and $c_{10}$ can be found in Ref.~\cite{Altmannshofer:2021qrr}.

Assuming the new physics appears at a scale above the weak scale $\Lambda$, 
the SMEFT Lagrangian with dimension 6 operators ($\mathcal{O}_i$) is defined as
\begin{eqnarray}
  \mathcal{L}_{SMEFT} &=& \mathcal{L}_{SM} + \frac{1}{\Lambda^2}\sum_{i}C_i\mathcal{O}_i,  
\end{eqnarray}
where $C_i$ are called Wilson Coefficients.

A complete set of non-redundant dimension 6 operators has been derived in Ref.~\cite{Grzadkowski:2010es}, so-called Warsaw basis. 
The most relevant operators in this paper are
\begin{eqnarray}
   \mathcal{O}^{(1)}_{lq}(p,r,s,t) &=& (\bar{l}_p\gamma^\mu l_r)(\bar{q}_s\gamma_\mu q_{t}),  \label{O1lq} \\ 
   \mathcal{O}^{(3)}_{lq}(p,r,s,t) &=& (\bar{l}_p\gamma^\mu\tau^I l_r)(\bar{q}_s\gamma_\mu\tau^I q_{t}),  \label{O3lq} \\ 
   \mathcal{O}_{qe}(p,r,s,t) &=& (\bar{q}_p\gamma^\mu q_r)(\bar{e}_s\gamma_\mu e_t), \label{Oqe} \\
   \mathcal{O}_{ld}(p,r,s,t) &=& (\bar{l}_p\gamma^\mu l_r)(\bar{d}_s\gamma_\mu d_t), \label{Old} \\
   \mathcal{O}_{ed}(p,r,s,t) &=& (\bar{e}_p\gamma^\mu e_r)(\bar{d}_s\gamma_\mu d_t), \label{Oed} 
\end{eqnarray}
where $q$, $l$, $e$ are the left-handed quark doublet, left-handed lepton doublet, and right-handed lepton singlet, respectively. 
$p,r,s,t$ are generation indices of quark or lepton.

When the electroweak symmetry breaking occurs, the SM heavy particles (top, Higgs,  $W^\pm$,  $Z$) are integrated out, and the SMEFT should be matched to LEFT. 
The full matching conditions at the tree level have been derived by Ref.~\cite{Jenkins:2017jig}.
In this paper, we only consider the operators with flavor indices $(p,r,s,t)=(2,2,2,3)$ or $(p,r,s,t)=(2,3,2,2)$, so the matching conditions are simplified to 
\begin{eqnarray}
    L^{V,LL}_{ed}(2,2,2,3) &=& \frac{v^2}{\Lambda^2}\left[C^{(1)}_{lq}(2,2,2,3)+C^{(3)}_{lq}(2,2,2,3)\right],  \label{match:VLLed}\\
    L^{V,LR}_{de}(2,3,2,2)&=& \frac{v^2}{\Lambda^2}C_{qe}(2,3,2,2), \label{match:VLRde} \\
    L^{V,LR}_{ed}(2,2,2,3)&=& \frac{v^2}{\Lambda^2}C_{ld}(2,2,2,3), \label{match:VLRed} \\
    L^{V,RR}_{ed}(2,2,2,3)&=& \frac{v^2}{\Lambda^2}C_{ed}(2,2,2,3). \label{match:VRRed}
\end{eqnarray}
The related renormalization group equations for the SMEFT and LEFT can be found in Appendix \ref{smeftrge} and Appendix \ref{leftrge}, respectively.
In this work, we do not distinguish the left-handed and right-handed contributions. 
The constraints on $C^{(1)}_{lq}+C^{(3)}_{lq}$ and $C_{qe}$, 
which are corresponding to $c_{9}$ and $c_{10}$, 
can also be applied to $C_{ld}$ and $C_{ed}$, 
which are corresponding to $c^{\prime}_{9}$ and $c^{\prime}_{10}$. Especially since  $O^{\prime}_{9}$ and $O^{\prime}_{10}$ are only related to the right-handed part of down type quarks, they do not contribute to $\mu^+\mu^- \rightarrow tc$ directly at higher energy
 in our main study.
 From now on, we only consider three operators: 
$\mathcal{O}^{(1)}_{lq}$, $\mathcal{O}^{(3)}_{lq}$ and $\mathcal{O}_{qe}$.

In Table.~\ref{table:LVLLed}, we list the runnings of three benchmark points: 
1) BP1, $c_{9}=-1.0, c_{10}=-0.1$, as an example of general new physics scenario, 
2) BP2, $c_{9}=c_{10}=0.25$, as an example that $C_{qe}$ is the main contribution at scale beyond $M_Z$,
3) BP3, $c_{9}=-c_{10}=-0.39$, as an example that $C^{(1)}_{lq}$ and $C^{(3)}_{lq}$ are the main contributions at scale beyond $M_Z$. 
These benchmark points are allowed in the analysis of Ref.~\cite{Ciuchini:2022wbq}, 
which has considered the newest LHCb data.
The BP3 is the best fit in Ref.~\cite{Altmannshofer:2021qrr}. 
It is observed that the operator mixings induced by the RGE running have no large effects on the size of Wilson coefficients of the SMEFT at high energy machines.

\begin{center}
\begin{table}
  \begin{center}
  \begin{tabular}{|l|c|c|c|}
  \hline
  & BP1 & BP2 ($c_9=c_{10}$)& BP3($c_9=-c_{10}$) \\
  \hline
$c_9$  &      $-1.00$        &     $0.25$         &       $-0.39$         \\
$c_{10}$  &      $-0.10$     &     $0.25$       &       $+0.39$       \\
  \hline  
  $L^{V,LL}_{ed}(m_B)$                   &   $-2.13\times 10^{-5}$   &  $0.00$                &  $-1.85\times 10^{-5}$    \\
  $L^{V,LR}_{de}(m_B)$                   &   $-2.60\times 10^{-5}$   &  $1.18\times 10^{-5}$  &  $0.00$                   \\ 
\hline
  $L^{V,LL}_{ed}(m_Z)$                   &   $-2.16\times 10^{-5}$   &  $2.83\times 10^{-8}$  &  $-1.86\times 10^{-5}$       \\
  $L^{V,LR}_{de}(m_Z)$                   &   $-2.60\times 10^{-5}$   &  $1.18\times 10^{-5}$  &  $-4.41\times 10^{-8}$    
\\
  \hline
  $(C^{(1)}_{lq}+C^{(3)}_{lq})(m_Z)$     &   $-3.56\times 10^{-2}$   &  $4.67\times 10^{-5}$  &  $-3.10\times 10^{-2}$    \\
  $C_{qe}(m_Z)$                          &   $-4.29\times 10^{-2}$   &  $1.95\times 10^{-2}$  &  $-7.28\times 10^{-5}$    \\
  \hline
  $(C^{(1)}_{lq}+C^{(3)}_{lq})(\Lambda)$ &   $-3.75\times 10^{-2}$   &  $1.03\times 10^{-4}$  &  $-3.20\times 10^{-2}$    \\                    
  $C_{qe}(\Lambda)$                      &   $-4.42\times 10^{-2}$   &  $2.00\times 10^{-2}$  &  $-2.02\times 10^{-4}$   \\
  \hline
  \end{tabular}
  \end{center}
  \caption{The coefficients $L^{V,LL}_{ed}$, $L^{V,LR}_{de}$, $C^{(1)}_{lq}+C^{(3)}_{lq}$ and $C_{qe}$ at scale $m_B=5$ GeV, $m_Z=91.19$ GeV and $\Lambda=10$ TeV are listed, with different input of $c_9$ and $c_{10}$. Here, we assume $C^{(1)}_{lq}=L^{V,LL}_{ed}\times \Lambda^2/v^2, C^{(3)}_{lq}=0$. \label{table:LVLLed}}
\end{table}
\end{center}

\section{The Matching Conditions of New Physics}
\label{Sec:NP}

At muon colliders, the operators we input are Eq.(\ref{O1lq}) to Eq. (\ref{Oqe}).
 In massless limit,\footnote{For simplicity, in this section we work under the massless limit. Nevertheless, for the numerical results discussed in later sections, full mass dependence is always included.} the differential cross sections for $\mu^+\mu^-\to tc$ and  $\mu^+\mu^-\to bs$ are:
 \begin{align}
     \frac{\ud\sigma(\mu^+\mu^-\to X)}{\ud\cos\theta}=\frac{3s}{256\pi \Lambda^4}\left[|C_{LL}^{X}|^2(1+\cos\theta)^2+|C_{qe}|^2(1-\cos\theta)^2\right]
 \end{align}
 where
 \begin{align}
 C_{LL}^{bs}=C_{lq}^{(1)}+C_{lq}^{(3)},C_{LL}^{tc}=C_{lq}^{(1)}-C_{lq}^{(3)}
 \end{align}

We note that the left-handed operators($C_{LL}^{X}$) and right-handed operator($C_{qe}$) have different $\theta$ dependence, hence we expect that with flavor tagging and charge identification, they can be distinguished with differential cross section.
For simplicity, in this work, we consider only inclusive cross section, which can be obtained by integrating over $\cos\theta$, given by:
\begin{align}
    \sigma(\mu^+\mu^-\to X)=\frac{1}{32\pi\Lambda^4}s(|C_{LL}^{X}|^2+|C_{qe}|^2)
\end{align}

In a general new physical model,
all three operators $C_{lq}^{(1)}$, $C_{lq}^{(3)}$ and $C_{qe}$ can be non-zero,
and thus both processes $\mu^+\mu^-\to tc$ and $\mu^+\mu^-\to bs$ receives new physical contribution,
to be measured in future muon colliders.
We note that for $\mu^+\mu^-\to bs$, the corresponding Wilson coefficients $C_{LL}^{bs}$ and $C_{qe}$ are directly in charge of $b\to s\mu^+\mu^-$ transition in $B$-physics,
and hence constrained by those measurements.
On the other hand, $C_{LL}^{tc}$ is unconstrained.
As a general argument, the underlying new physical which induces $C_{LL}^{bs}$ or equivalently $C_{lq}^{(1)},C_{lq}^{(3)}$,
would induce also $C_{LL}^{tc}$ with size at similar order.

Therefore, we expect that the cross-section of $\mu^+\mu^-\to tc$ is comparable to $\mu^+\mu^-\to bs$,
and as we will show in Section \ref{Sec:result}, due to the presence of top quark in the final state of $\mu^+\mu^-\to tc$,
it is easier to measure than $\mu^+\mu^-\to bs$.
To be more specific, below we discuss four types of new physics models labeled as Model I-IV, 
where the relations between $C_{LL}^{tc}$ and $C_{LL}^{bs}$ are given,
and later in Section \ref{Sec:result} we will show how to distinguish them by measuring both $\mu^+\mu^-\to tc$ and $\mu^+\mu^-\to bs$.

\begin{enumerate}
    \item[Model I]   A $Z^{\prime}$ model with flavor symmetry $U(1)_{L_{\mu}-L_{\tau}}$.
        In this model, the $L_{\mu}-L_{\tau}$ is promoted into a $U(1)$ gauge symmetry,
      with a massive gauge boson $Z^{\prime}$.
    Originally it was proposed for muon $g-2$ anomaly \cite{Baek:2001kca,Allanach:2015gkd} and neutrino mixing \cite{Ma:2001md}.
    Later it is realized that the coupling to quarks can be described by high dimensional operators, which can be generated through new heavy quarks \cite{Fox:2011qd}.
    Such interaction can induce flavor violation \cite{Altmannshofer:2014cfa,Crivellin:2015mga,Crivellin:2015lwa}.
    Since the effective interaction between $bs\mu\mu$($tc\mu\mu$) is mediated by an $s$-channel $Z^{\prime}$, an electroweak singlet,
    clearly we have $C_{lq}^{(3)}=0$, hence $C_{LL}^{bs}=C_{LL}^{tc}=C_{lq}^{(1)}$.
    On the other hand, in this kind of models $C_{qe}$ is independent from $C_{lq}^{(1)}$.
    Regardless of the actual value of $C_{qe}$ and $C_{lq}^{(1)}$,
    we have $\sigma(tc)\sim\sigma(bs)$.

    \item[Model II] A scalar triplet leptoquark $S_3$ model.
        In this model, the new physics is mediated by a heavy scalar leptoquark, which belongs to $SU(2)_L$ triplet.
        The corresponding Lagrangian can be written as \cite{Dorsner:2016wpm}
        \begin{align}
            \mathcal{L}_{NP}=\sum_{i,j}\lambda_{ij}\bar{Q}_i^c(i\tau^2)\tau^{I}L_jS_3^{I}+\textrm{h.c.}
        \end{align}
        In this model, both $C_{lq}^{(1)}$ and $C_{lq}^{(3)}$ can be generated at tree-level, which is given by \cite{Gherardi:2020det}
        \begin{align}
            \frac{1}{\Lambda^2}[C_{lq}^{(1)}]_{prst}=&3\frac{\lambda_{sp}^{*}\lambda_{tr}}{4M^2}\\
            \frac{1}{\Lambda^2}[C_{lq}^{(3)}]_{prst}=&\frac{\lambda_{sp}^{*}\lambda_{tr}}{4M^2}.
        \end{align}
        where $M$ is the mass of the leptoquark $S_3$.
        On the other hand, $C_{qe}$ is zero at tree-level, and it can only be generated through loop corrections, and hence we expect $C_{qe}\ll C_{lq}^{(1)},C_{lq}^{(3)}$.
        Therefore, we have $C_{LL}^{bs}\sim2C_{LL}^{tc}$,
        and $\sigma(tc)\sim\frac{1}{4}\sigma(bs)$

    \item[Model III] A scalar singlet leptoquark $S_1$ model.
        In this model, new physics is mediated by a heavy scalar leptoquark, which belongs to $SU(2)_{L}$ singlet.
        There are three possible hypercharge assignments,
        and we consider the case $Y=\frac{1}{3}$ here.
        The corresponding Lagrangian is
        \begin{align}
            \mathcal{L}_{NP}=\sum_{i,j}\lambda_{ij}\bar{Q}^c_i(i\tau_2)L_jS_1+\textrm{h.c.}
        \end{align}
        At the tree level, the relevant Wilson coefficients are
        \begin{align}
            \frac{1}{\Lambda^2}[C_{lq}^{(1),\textrm{tree}}]_{prst}=\frac{\lambda_{sp}^{L*}\lambda_{tr}^{L}}{4M^2}\\
            \frac{1}{\Lambda^2}[C_{lq}^{(3),\textrm{tree}}]_{prst}=-\frac{\lambda_{sp}^{L*}\lambda_{tr}^{L}}{4M^2}
        \end{align}
        Interestingly, we can see that at tree-level $C_{LL}^{bs}=0$.
        The leading contribution to $C_{LL}^{bs}$ starts from one-loop level \cite{Bauer:2015knc}.
        Consequently, we expect that $C_{LL}^{bs}\ll C_{LL}^{tc}$,
        and hence $\sigma(tc)\gg C_{LL}^{bs}$.
        For the experiment bounds($C_9=-1$ benchmark point), we get
        \begin{align}
            \sum_{i}|\lambda_{u_i\mu}|^2\mathrm{Re}\frac{(\lambda\lambda^{\dag})_{bs}}{V_{tb}V_{ts}^{*}}-1.74|\lambda_{t\mu}|^2\sim 12.5\hat{M}^2
        \end{align}
        where $\hat{M}$ is the mass of the leptoquark in terms of unit TeV,
        and from $B_s-\bar{B}_s$ mixing we have
        \begin{align}
            \frac{(\lambda\lambda^{\dag})_{bs}}{V_{tb}V_{ts}^{*}}\sim(1.87+0.45i)\hat{M}
        \end{align}
        For perturbativity, we have $|\lambda_{u_i\mu}|^2<4\pi$, which yields
        \begin{align}
            M\lesssim 1.9\si{TeV}
        \end{align}
        Consequently, we expect that at the energy range of future muon colliders, a pair of the singlet scalars $S_1$  can be produced and be observed directly.
    \item[Model IV] A $U_1$ vector leptoquark model. It was considered as one possible UV completion of EFT in \cite{Calibbi:2015kma}, and examined in more detail in \cite{Barbieri:2015yvd,Calibbi:2017qbu,Blanke:2018sro,Buttazzo:2017ixm}.
        The Lagrangian is given by 
        \begin{align}
        \mathcal{L}_{NP}=-\frac{1}{2}U_{1,\mu\nu}^{\dag}U^{1,\mu\nu}+M_U^2U_{1,\mu}^{\dag}U_1^{\mu}+g_UU_{1,\mu}\lambda_{ij}\bar{Q}_{i}\gamma^{\mu}L_{j}+\textrm{h.c.}
        \end{align}
        The relevant Wilson coefficients are
        \begin{align}
            \frac{1}{\Lambda^2}[C_{lq}^{(1),\textrm{tree}}]_{prst}=g_U^2\frac{\lambda_{sp}^{L*}\lambda_{tr}^{L}}{2M^2}\\
            \frac{1}{\Lambda^2}[C_{lq}^{(3),\textrm{tree}}]_{prst}=g_U^2\frac{\lambda_{sp}^{L*}\lambda_{tr}^{L}}{2M^2}
        \end{align}
        Clearly, in this case we have $C_{LL}^{tc}\sim0,C_{LL}^{bs}\gg C_{LL}^{tc}$. Moreover, in this minimal setup we have $C_{qe}=0$, though it can be introduced through adding other particles and/or interactions.
\end{enumerate}
We'd like to note that the above are examples of simplest new physics models, where only one type of mediator is introduced. The Nature may be more complicated and contains several types of mediators, e.g. both scalar single and triple leptoquarks may be present\cite{Crivellin:2017zlb}.

\section{Signatute of $\mu^+\mu^-\to tc$ at future muon colliders }
\label{Sec:result}

\subsection{Cross sections of signal and main background processes}
\begin{figure}
  \centering
  \captionsetup[sub]{font=large}
  \begin{subfigure}[t]{0.45\textwidth}
     \includegraphics[width=\linewidth]{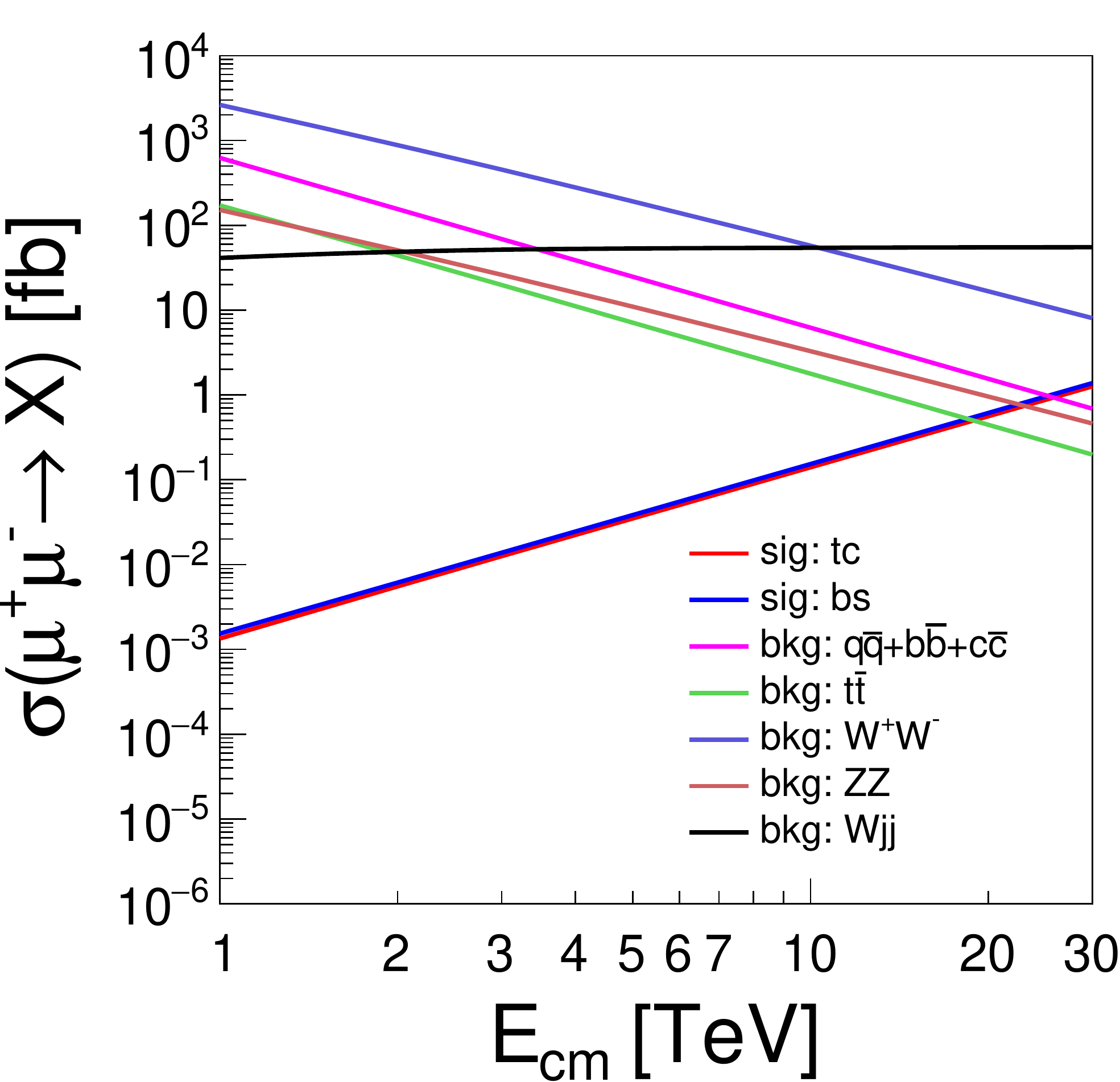}
     \caption{}
  \end{subfigure}
  \begin{subfigure}[t]{0.45\textwidth}
     \includegraphics[width=\linewidth]{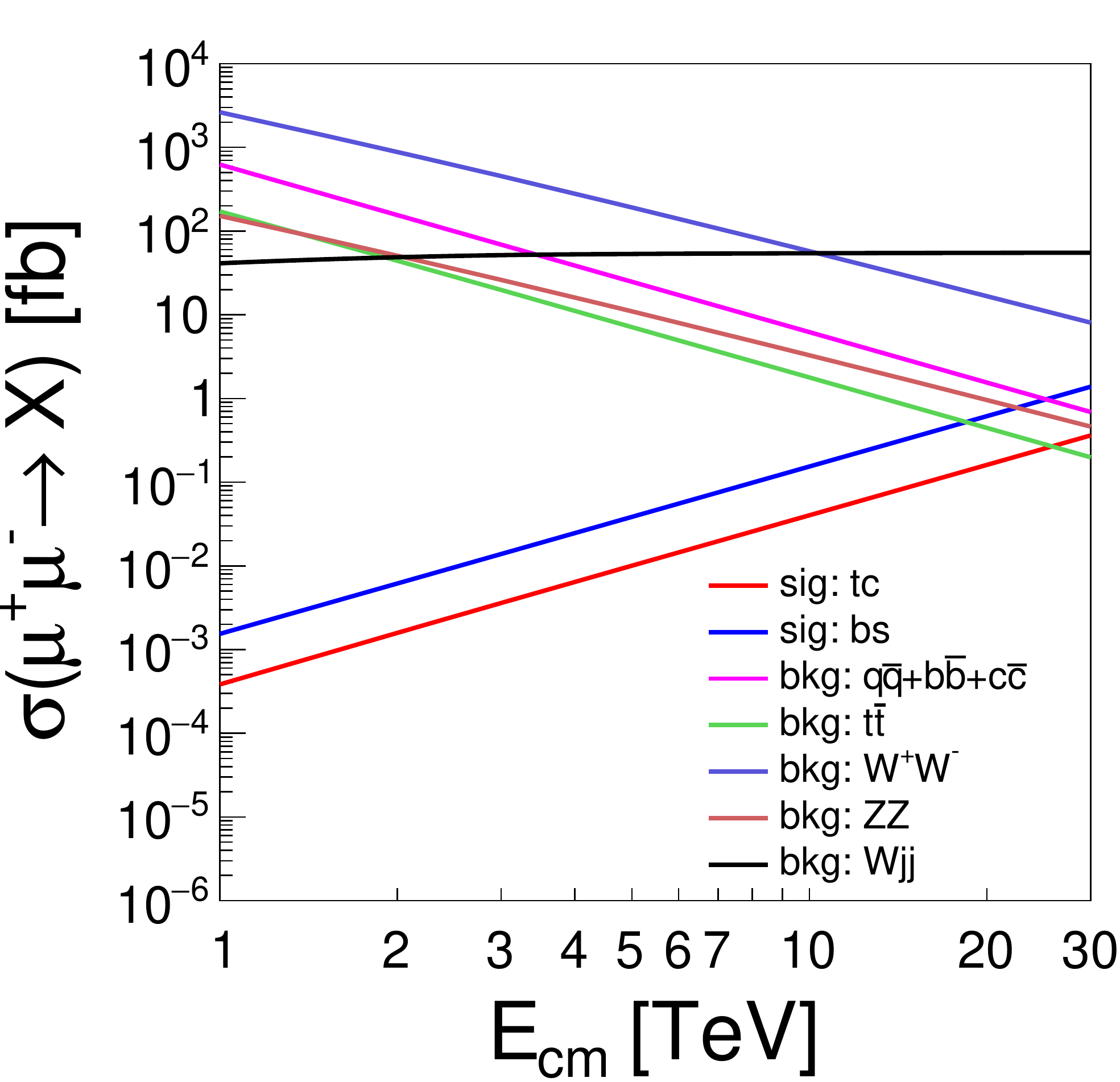}
     \caption{}
  \end{subfigure}
  \caption{The energy dependencies of the cross sections of $\mu^+\mu^-\to X$ are displayed, with generator cuts: $p_T(j)>20$ GeV, $|\eta(j)|<5.0$ and $\Delta R(j,j)>0.4$. For the signal, $X=tc/bs$. In other cases, $X$ is the final state of the background. The BP3 with $c_{9}=-c_{10}=-0.39$ is shown here. Their values are evaluated below the cutoff $\Lambda=10$ TeV by our RGEs and matching conditions in the Appendix. Two models are considered: (a) Model I, $C^{(1)}_{lq}=L^{V,LL}_{ed}\times \Lambda^2/v^2$ and $C^{(3)}_{lq}=0$,  and (b) Model II, $C^{(1)}_{lq}=3C^{(3)}_{lq}=\frac{3}{4}L^{V,LL}_{ed}\times \Lambda^2/v^2$.\label{ecm}}
\end{figure}

To study the $\mu^+\mu^-\to tc$ process at future muon collider, 
we generate a UFO model with relevant operators by Feynrules~\cite{Alloul:2013bka}, 
and import it to \mgamc \cite{Alwall:2014hca}. 
We calculate the cross sections of signals and backgrounds from $E_{cm}=1$ TeV to $30$ TeV, 
with following generator cuts: jet transverse momentum $p_T(j)>20$ GeV, jet pseudorapidity $|\eta(j)|<5.0$, and angular distance between jets $\Delta R(j,j)>0.4$.
The energy dependencies of the cross sections are shown in Fig.~\ref{ecm}. 

For the $\mu^+\mu^-\to tc$ and $bs$ processes, we consider the BP3 as an example where $c_{9}=-c_{10}=-0.39$. The cross section of the main signal process $\mu^+ \mu^- \to tc$ derived from the 4-fermion interactions are proportional to collision $s^2$, 
and it increases with the increase of the collision energy, as shown in Fig.~\ref{ecm}. 
For Wilson coefficients $C^{1}_{lq}$ and $C^{3}_{lq}$, we consider their relations in Model I and II. 
As we expect, their cross sections have relation $\sigma(tc)\sim \sigma(bs)$ for Model I, and $\sigma(tc)\sim\frac{1}{4}\sigma(bs)$ for Model II. 

The main background processes include $b\bar{b}$, $c\bar{c}$, $q\bar{q}$ ($q=u,d,s$), $W^+ W^-$, $ZZ$, $t \bar{t}$, and $Wjj$ ($j=u,d,s,c$). 
These processes are from S-channel and decrease with the increase of the collision energy. 
It is noteworthy that the cross sections of diboson processes $W^+W^-$ can be much larger than those of di-quarks final states. 
The process $W^\pm jj$ is also included, which describes the weak final state radiation. 
A real $W$ boson is radiated from di-quark process, which introduces a $\ln^2(s/M^2_{W})$ term to the cross section~\cite{Bell:2010gi}.  
$W^\pm jj$ cross section is around $55$ fb and remains constant even when the collision energy increases. 
It is remarkable that the cross section of $W^\pm jj$ is much larger than those of other processes. 
It actually shows the impact of electroweak radiation. 
A more clever way should be doing analysis with full next-to-leading order corrections, 
but it is a huge challenge. 
On one hand, 
due to the non-abelian nature of electroweak symmetry, the vitual and real corrections do not fully cancel~\cite{Ciafaloni:2000df,Ciafaloni:2000rp}.
On the other hand, 
the W/Z mass suggests an IR scale $\sim 100$ GeV, and it is physical. 
Electroweak shower always ends at this scale and the electroweak ``partons'' do not ``hadronize''. 
Pythia8 only provides limited features to the electroweak shower~\cite{Christiansen:2014kba,Christiansen:2015jpa}. 
A summary of these challenges can be found in Ref.~\cite{Accettura:2023ked}.
As a tree-level analysis, we have to treat it as a separated process.
Therefore, suppressing the background events of $\mu^+ \mu^- \to W^\pm jj$ will be crucial and necessary.

Near the threshold, the cross section of all background processes is huge compared to the signal process. 
With the increase of collision energy, 
the signal cross section can even be larger than those of background processes if the collision energy is large enough (say $\sqrt{s}=30$ TeV) for some new physics models. 
Nonetheless, when $\sqrt{s}=10$ TeV, it might be still challenging to discover the signal events since the cross section of background processes is several orders larger than that of the signals processes.

\begin{figure}
  \centering
  \captionsetup[sub]{font=large}
  \begin{subfigure}[t]{0.45\textwidth}
     \includegraphics[width=\linewidth]{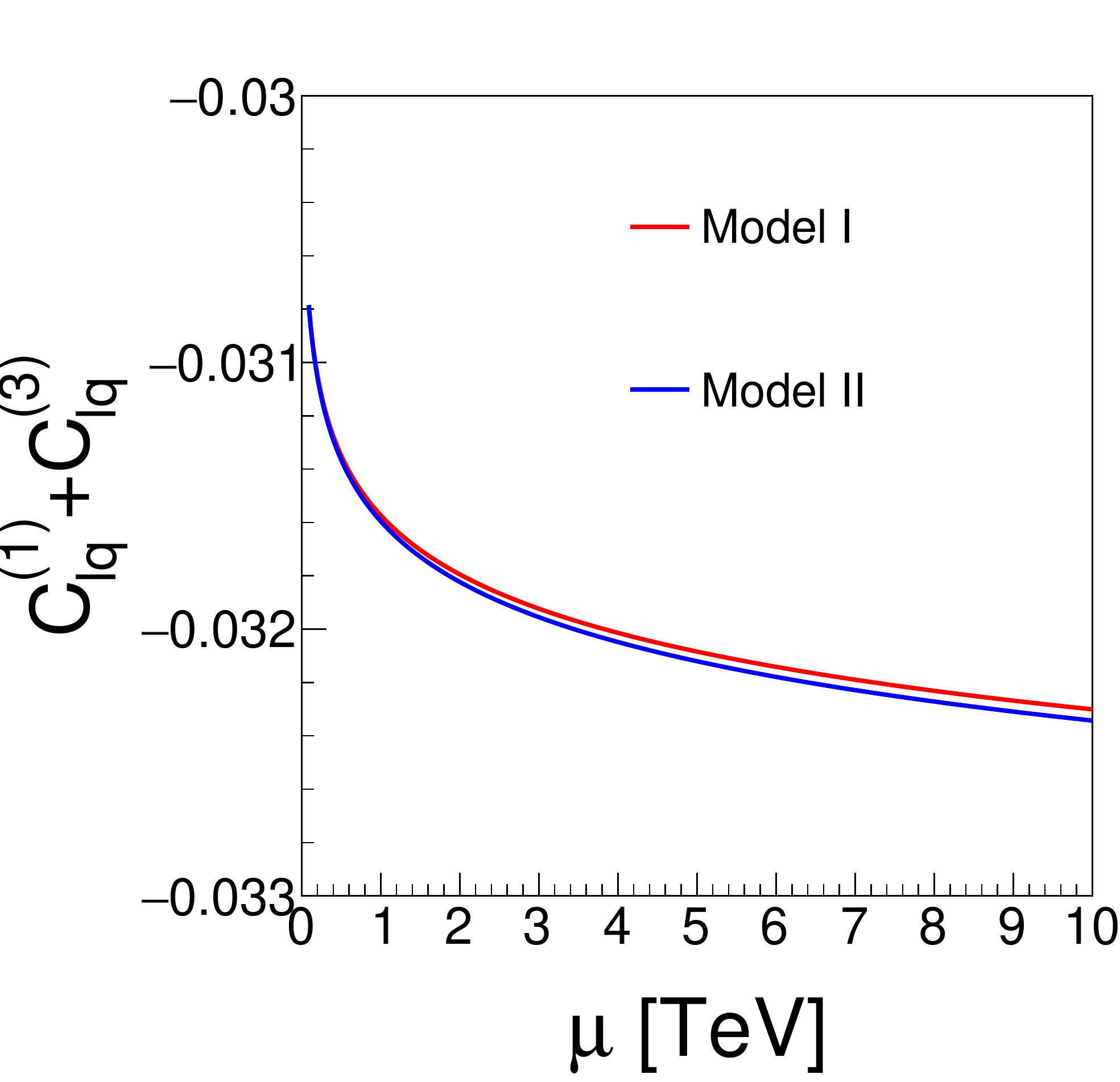}
     \caption{}
  \end{subfigure}
  \begin{subfigure}[t]{0.45\textwidth}
     \includegraphics[width=\linewidth]{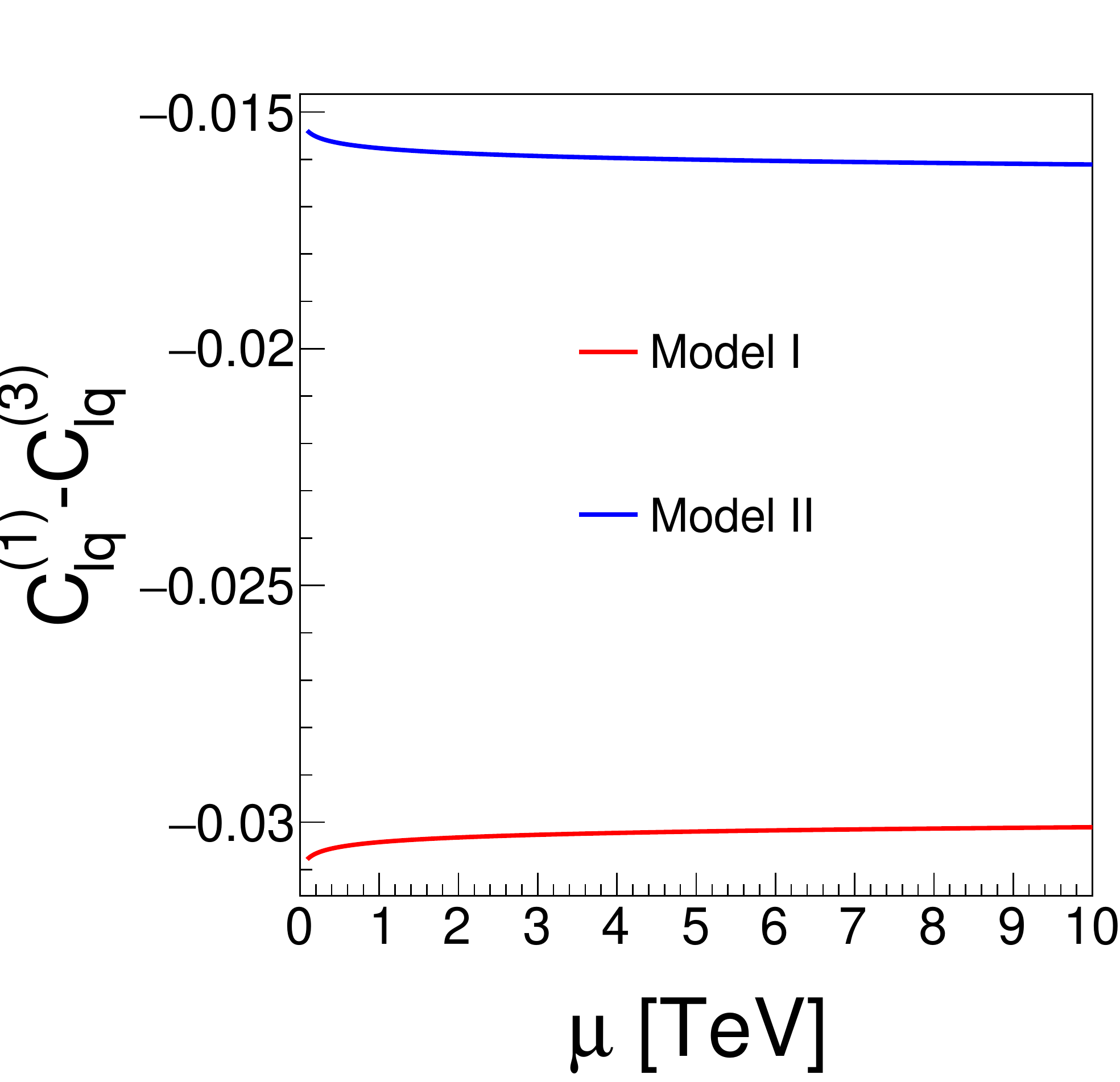}
     \caption{}
  \end{subfigure}
  \caption{The runnings of (a) $C^{(1)}_{lq}+C^{(3)}_{lq}$ and (b) $C^{(1)}_{lq}-C^{(3)}_{lq}$ from EW scale to $\Lambda=10$ TeV are displayed.  Two NP scenarios are considered: 1) Model I, $C^{(1)}_{lq}=L^{V,LL}_{ed}\Lambda^2/v^2$ and $C^{(3)}_{lq}=0$, and 2) Model II, $C^{(1)}_{lq}=3C^{(3)}_{lq}=\frac{3}{4}L^{V,LL}_{ed}\times \Lambda^2/v^2$.\label{rge:EWtoUV}}
\end{figure}

As we show in Fig.~\ref{ecm}, the process $\mu^+\mu^-\to tc$ can only be observed when $E_{cm}>10$ TeV. 
At such high energy, we can expect two energetic jets to be observed at the detector for either signal or background events. 

In Fig.~\ref{ecm}, the BP3 with $c_{9}=-c_{10}=-0.39$ given in Table \ref{table:LVLLed} is considered.
Their values are evaluated to the EW scale by our RGEs Eq.~\ref{beta:VLLed} and Eq.~\ref{beta:VLRde}. 
And then they are matched to SMEFT by Eq.~\ref{match:VLLed} and Eq.~\ref{match:VLRde}. 
From Eq.~\ref{match:VLLed}, we also know that the constraint from B physics can be only applied to the sum of $C^{(1)}_{lq}+C^{(3)}_{lq}$~\footnote{The generation indices are always $(p,r,s,t)=(2,2,2,3)$ for $C^{(1)}_{lq}$ and $C^{(3)}_{lq}$ and $(p,r,s,t)=(2,3,2,2)$ for $C_{qe}$ in this paper.}.  
So any new physics scenarios with $C^{(1)}_{lq}$ and $C^{(3)}_{lq}$ as free parameters cannot be constrained by the data from B physics experiments. 
In Fig.~\ref{rge:EWtoUV}, we show the running of $C^{(1)}_{lq}+C^{(3)}_{lq}$ and $C^{(1)}_{lq}-C^{(3)}_{lq}$ from EW scale to new physics scale at $\Lambda=10$ TeV for the BP3 case with $c_{9}=-c_{10}=-0.39$.
We have considered two simple cases at EW scale ($M_Z$): 1)  Model I with $C^{(1)}_{lq}=L^{V,LL}_{ed} \frac{\Lambda^2}{v^2}$ and $C^{(3)}_{lq} = 0$; 2) Model II with $C^{(1)}_{lq}=3 C^{(3)}_{lq}=\frac{3}{4} L^{V,LL}_{ed} \frac{\Lambda^2}{v^2}$.
Considering the experimental uncertainties, it is challenging to distinguish these two cases from the effects of RGE running of only $C^{(1)}_{lq}+C^{(3)}_{lq}$. 

It is noteworthy that although these two NP cases, i.e. Model I and Model II, produce the same $c_9$ and $c_{10}$ at $\mu=M_B$, their high energy behaviors are different from each other. 

\subsection{Jet Level Analysis}

It should be mentioned that for a muon collider, 
the collision environment is relatively clean and there are no serious pileups and underlying events that occurred at a hadron collider, 
but there exist beam-induced backgrounds which arise from the muon decay. 
By using the jet grooming techniques, 
it is possible to suppress such beam-induced backgrounds \cite{Collamati:2021sbv}. 
 We will neglect the beam-induced backgrounds in our study.

Meanwhile, since it is more and more realistic to use the particle flow method to measure the jet energy, 
which can reduce the uncertainty of jet energy down to $5\%-20\%$ \cite{Nachman:2022emq}. 
 The particle flow method can also help us to distinguish a light jet, 
a W/Z boson jet, and a top jet: a W/Z/top jet can have more charged tracks in a detector compared with a light jet.

To further investigate the kinematic features of jets from signal processes, 
we input events generated with BP3 and Model I to PYTHIA8~\cite{Bierlich:2022pfr} for parton shower and hadronization, 
and FastJet~\cite{Cacciari:2011ma} to reconstruct jets.  
It is expected that jet algorithms developed for electron-positron colliders can also be applied well to muon colliders. 
Therefore, we use the generalized $k_t$ algorithm for $e^+e^-$ collisions,
which is extended from a simple $k_t$ algorithm~\cite{Catani:1991hj}.
This algorithm defines two distances:
\begin{eqnarray}
  d_{ij} &=& \min(E^{2p}_i,E^{2p}_j)\frac{1-\cos\theta_{ij}}{1-\cos{R}}, \label{eq:genkt1}\\
  d_{iB} &=& E^{2p}_i, \label{eq:genkt2}
\end{eqnarray}
where $p$ and $R$ are inputs from user. 
If a $d_{ij}$ is smallest then particle $i$ and $j$ are recombined, 
while $d_{iB}$ is smallest then $i$ is called an ``inclusive jet''.
In this context, we choose $p=1$.

It should be pointed out that the denominator $(1-\cos{R})$ is replaced by $(3+\cos{R})$ while choosing $\pi<R<3\pi$ in FastJet.
In this case, $d_{iB}$ is always larger than $d_{ij}$ so that only one inclusive jet can be found.  
If we also choose $p=1$, the generalized $k_t$ algorithm is identical to the original $k_t$ algorithm~\cite{Catani:1991hj}, 
which only has a single distance:
\begin{eqnarray}
  d_{ij} &=& \min(E^{2}_i,E^{2}_j)\left(1-\cos\theta_{ij}\right), \label{eq:eekt}
\end{eqnarray}
and one can extract ``exclusive jet'' only. 
For the high energy muon collider, muon beams may radiate energetic particles which can have large angles to the beams. 
Such kinds of particles are included in the exclusive jets. 
So the original algorithm is not sufficient in our context.

Below we will demonstrate a case study with the collision energy $\sqrt{s}=10$ TeV at the jet level analysis, 
with the BP3 and Model I as theoretical input. 
We demand all signal and background events have hadronic decays and neglect those semi-leptonic and pure leptonic final states. 

Since the dominant background processes are $WW (ZZ)$, $tt$, 
and $jj$ which can have more than two energetic jets in the final state, 
in order to avoid the combinatoric issues, it is better to have less number of jets in the final state. 
For example, when $R=0.05$ and $E_j>100$ GeV,
the number of jets for a signal event can be much more than $10$, which is difficult to analyze. 
Then appropriate jet parameters, like cone parameter and energy cut, 
are crucial for the jet numbers and the preselection of signal events. 
Due to the fact that both $W/Z$ bosons and top quarks are highly boosted in the final state 
(each of them carries an energy 5 TeV), 
it is crucial to set the value of jet parameter so as to capture the whole decay products of $W/Z$ bosons and top quarks. Therefore, in this study, we will focus on the boosted events in our analysis. Considering that the future ECAL (HCAL) sub-detectors can have a granularity good enough,  there is no doubt that a future detector of a muon collider is able to resolve the substructure of such a highly boosted top jet. In principle, a future detector of a muon collider should be very similar to the detectors of CEPC and ILC.

The optimization of parameters $p$ and $R$ is a complicated task.
As a first attempt, we choose $p=1$ so it works similarly to a $k_t$ algorithm for hadron collider~\cite{Catani:1993hr,Ellis:1993tq}. 
For the $t\bar{c}(\bar{t}c)$ final state, we hope to find a $R$ parameter to reconstruct a heavy jet around top mass and a light jet.
We have scanned the $R$ parameter from $0.05$ to $0.15$ and have found that $R=0.10$ can satisfy this requirement. 

\begin{figure}[htbp]
  \centering
  \captionsetup[sub]{font=large}
  \begin{subfigure}[t]{0.45\textwidth}
     \includegraphics[width=\linewidth]{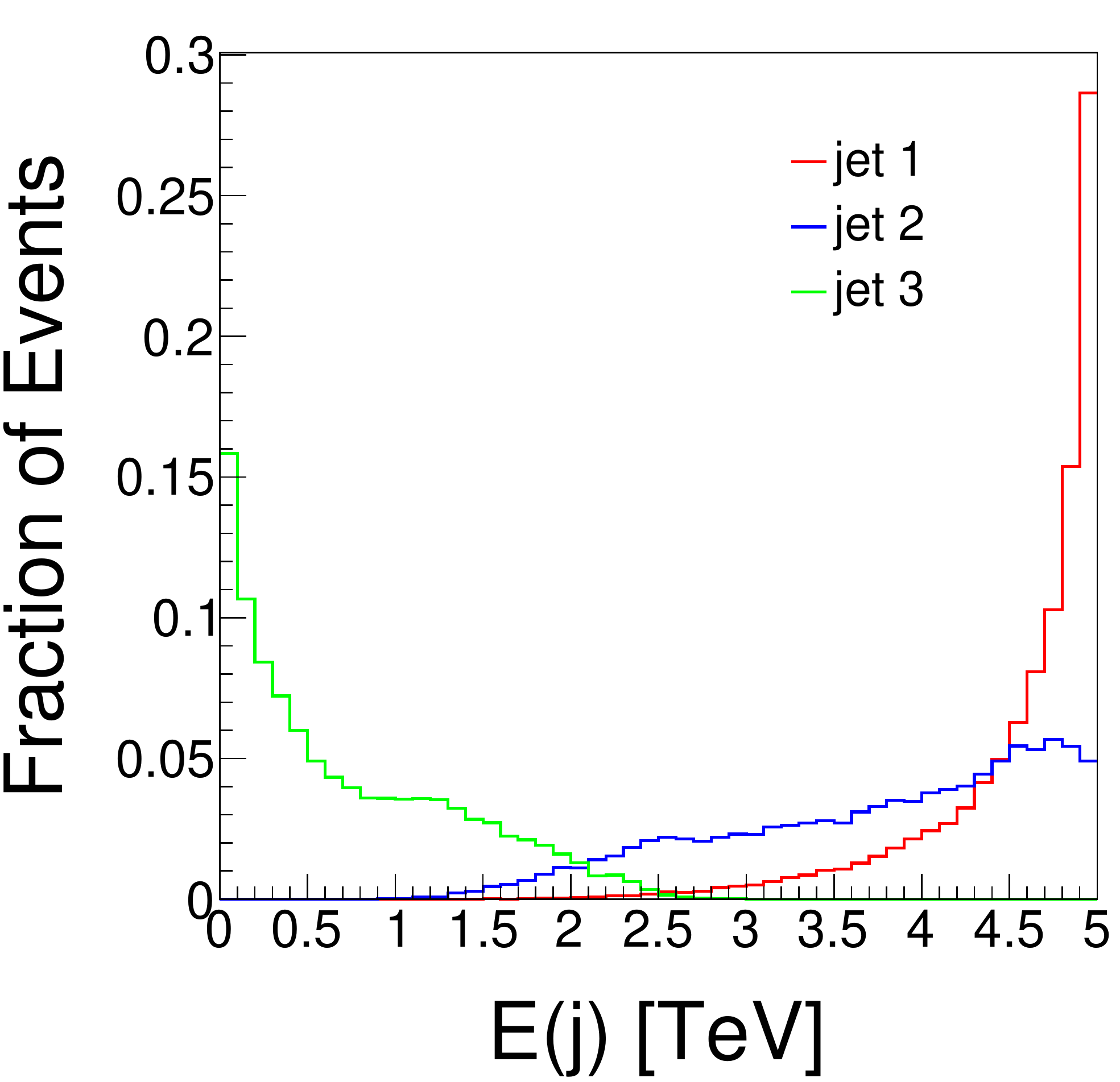}
     \caption{}
     \label{subfig:ej_tc}
  \end{subfigure}
  \begin{subfigure}[t]{0.45\textwidth}
     \includegraphics[width=\linewidth]{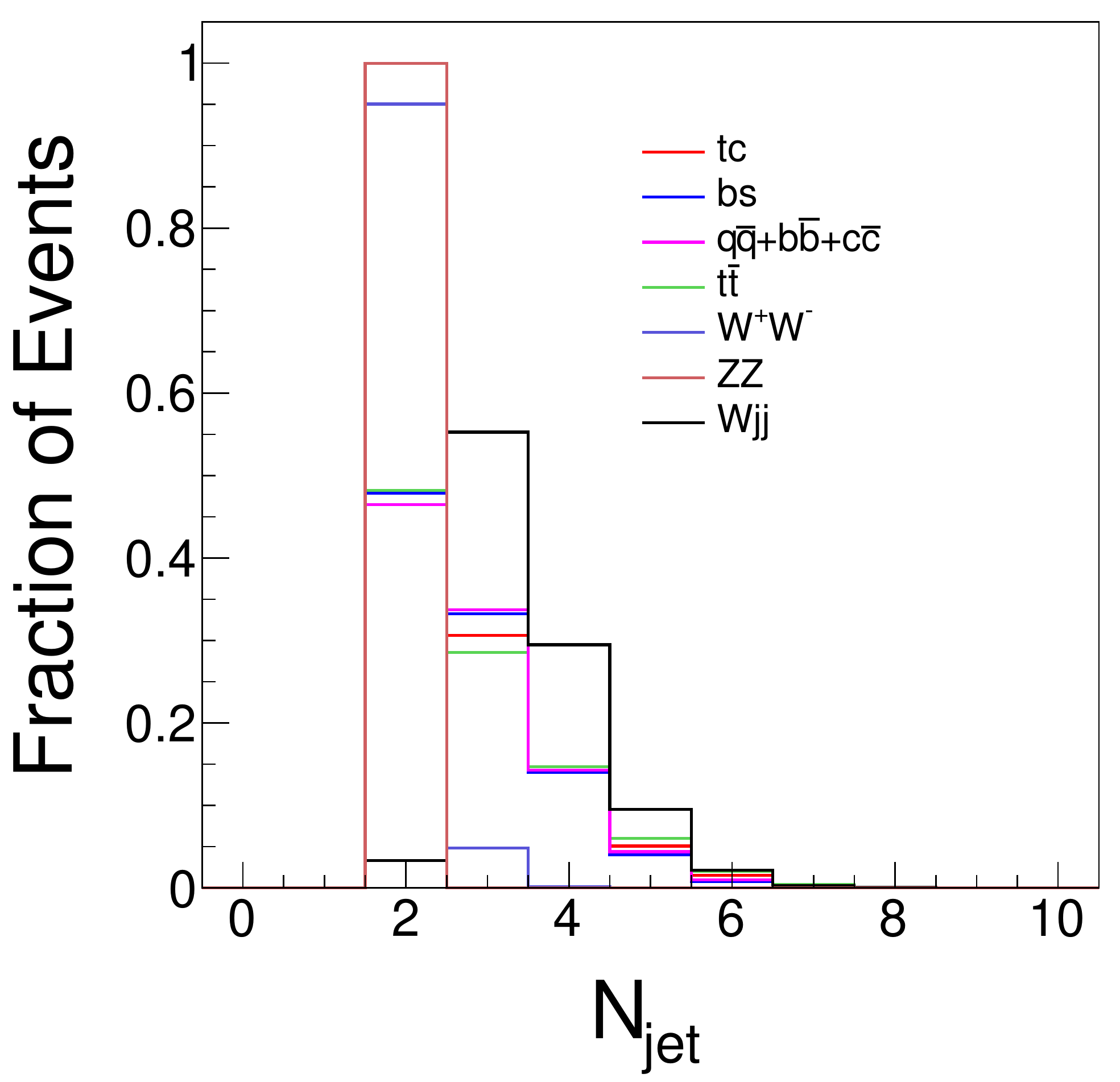}
     \caption{}
     \label{subfig:njets}
  \end{subfigure}
  \caption{(a) The jet energy distributions of process $\mu^+\mu^-\to tc$ are displayed, where jets are sorted by energy. (b) The number of jets after implementing cut $E(j)>500$ GeV.}\label{fig:base}
\end{figure}

In Fig.~\ref{subfig:ej_tc}, we display the energy distribution of the leading three jets in $t\bar{c}(\bar{t}c)$ final states with $R=0.10$ and $p=1$. 
These jets are sorted by energy.  Obviously, the first two jets have energy around $E_{cm}/2$, 
which means that they probably originated from a hard process.
We can also observe that the energy of the 3rd jet can reach several hundred GeV even TeV levels.
In such a high-energy collider, parton shower can radiate some high-energy particles and can be detected as a hard jet. To reduce these radiations, we implement a cut $E(j)>500$ GeV to jets for each event. 
After this cut, we plot the number of jets in Fig.~\ref{subfig:njets}. 
As we can see, the peak is $N_{jets}=2$ for two final state processes, and the peak for $W^{\pm}jj$ background is 3. 
Therefore to demand the number of jets $N_j=2$ for each event can heavily suppress the background events of $\mu^+ \mu^- \to W^{\pm}jj$.

\begin{figure}[htbp]
  \centering
  \captionsetup[sub]{font=large}
  \begin{subfigure}[t]{0.45\textwidth}
     \includegraphics[width=\linewidth]{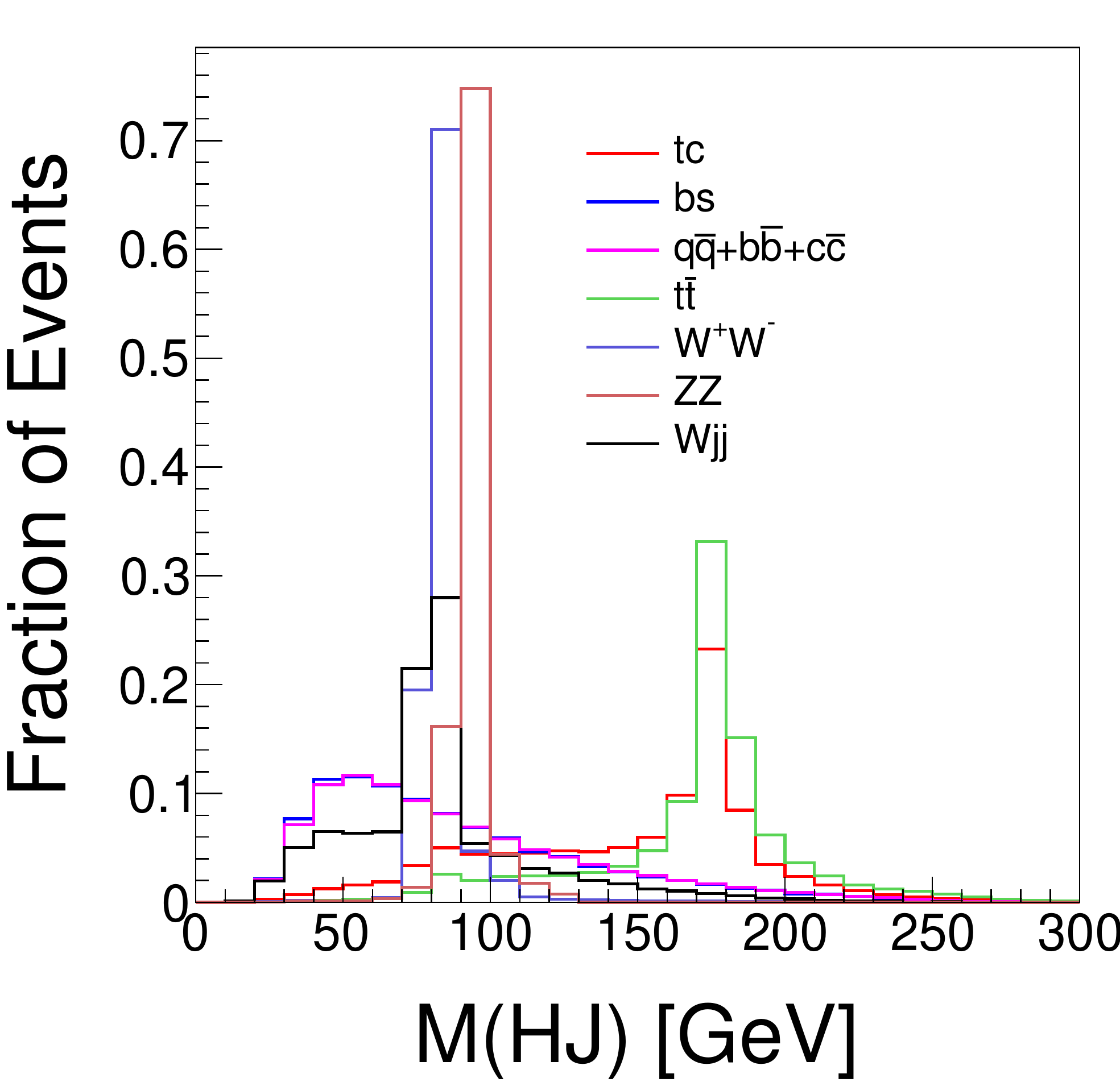}
     \caption{}
     \label{fig:mj1}
  \end{subfigure}
  \begin{subfigure}[t]{0.45\textwidth}
     \includegraphics[width=\linewidth]{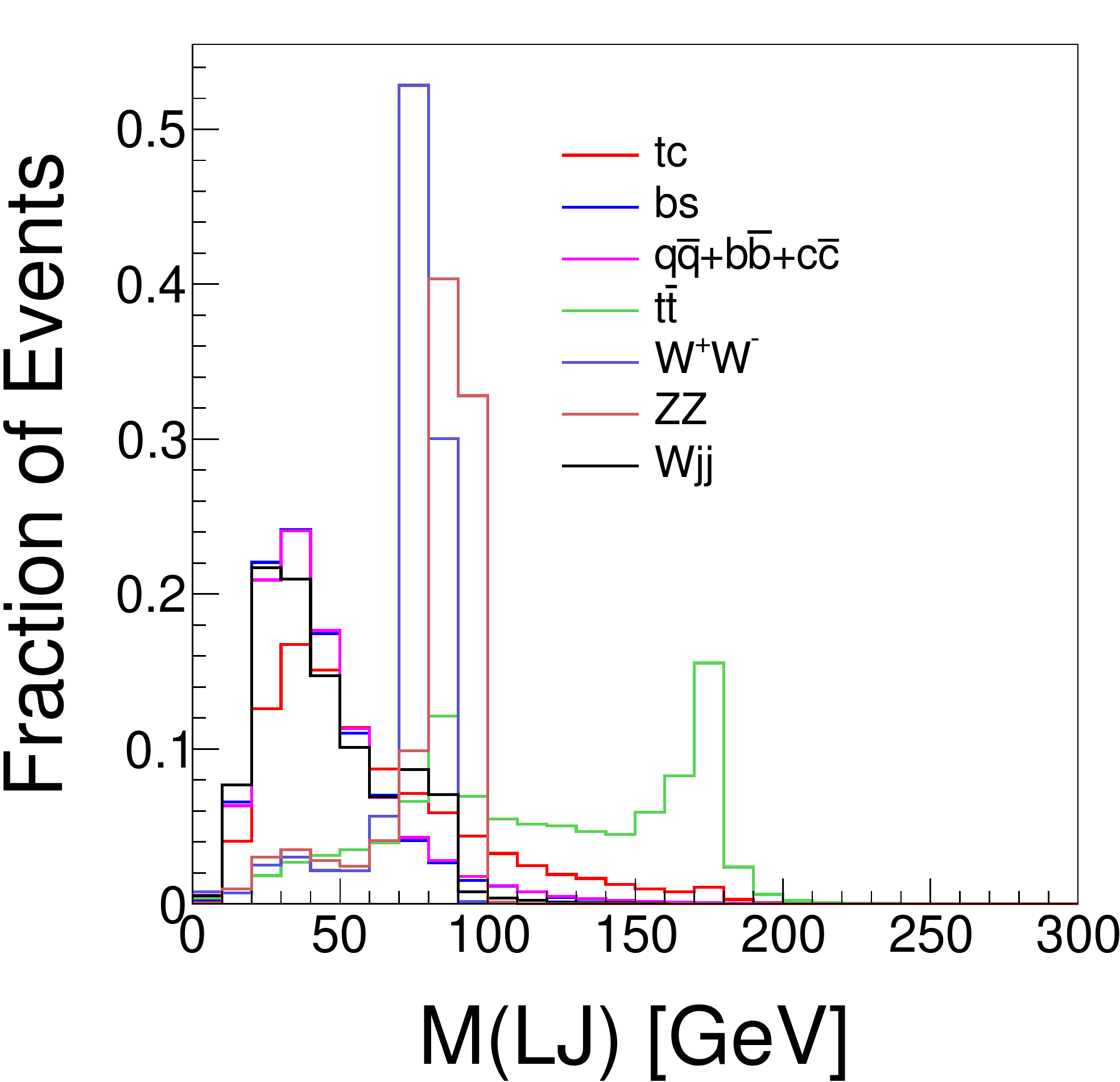}
     \caption{}
     \label{fig:mj2}
  \end{subfigure}
  \caption{The invariant mass of (a) the heavier jet and (b) the lighter jet in signal and backgrounds are displayed.}\label{fig:mj1mj2}
\end{figure}

Fig.~\ref{fig:mj1mj2} shows the invariant masses of the two most energetic jets of signal and background events. 
Here, we label the heavier jet as HJ in Fig.~\ref{fig:mj1} and the lighter one as LJ in Fig.~\ref{fig:mj2}.
As we expect, HJ has a peak around the top mass and the distribution of LJ is flat for the signal.
For the backgrounds with heavy particles ($t$, $W$, and $Z$), we also observe peaks around their masses. 
It is because these particles are highly boosted in such a high energy machine. 
The reconstruction of massive jets at ATLAS can be found in Ref.~\cite{ATLAS:2020gwe}.
It is shown that the $W$ and top jets mass resolution can reach $10\%$, 
with suitable reconstruction scheme. 
We can expect that this resolution can reach $1\sim 2\%$ at future muon collider, 
so applying cuts on jet invariant mass is helpful to extract the $tc$ signal.

B-tagging and C-tagging techniques can help with signal and background separation. 
As shown in \cite{ATLAS:2019bwq}, the larger the transverse momentum, the better the B tagging efficiency. 
Therefore we expect a higher B-tagging efficiency for a 10 TeV collision. 
Although B mesons and D mesons have similar proper times at $10^{-12}$ seconds, their masses are different. 
If C-tagging techniques \cite{ATLAS:2021cxe} can be applied in our analysis, we expect better results can be obtained. 

Since we only consider hadronic decay, the top quark should decay to $b$ quark and $W$ boson, 
and $W$ further decays to light quarks. 
So the HJ of a signal event should also be tagged as a $b$-jet in a future detector. 
To consider the $b$-tagging effects, we track all decayed products of $b$-hadrons after the hadronization. 
If the constituents of a jet have a b-hadron, we can label this jet as a true $b$-jet. 
The same procedure can be applied to labeling a $c$-jet. 

\begin{figure}[htbp]
  \centering
  \captionsetup[sub]{font=large}
  \begin{subfigure}[t]{0.45\textwidth}
     \includegraphics[width=\linewidth]{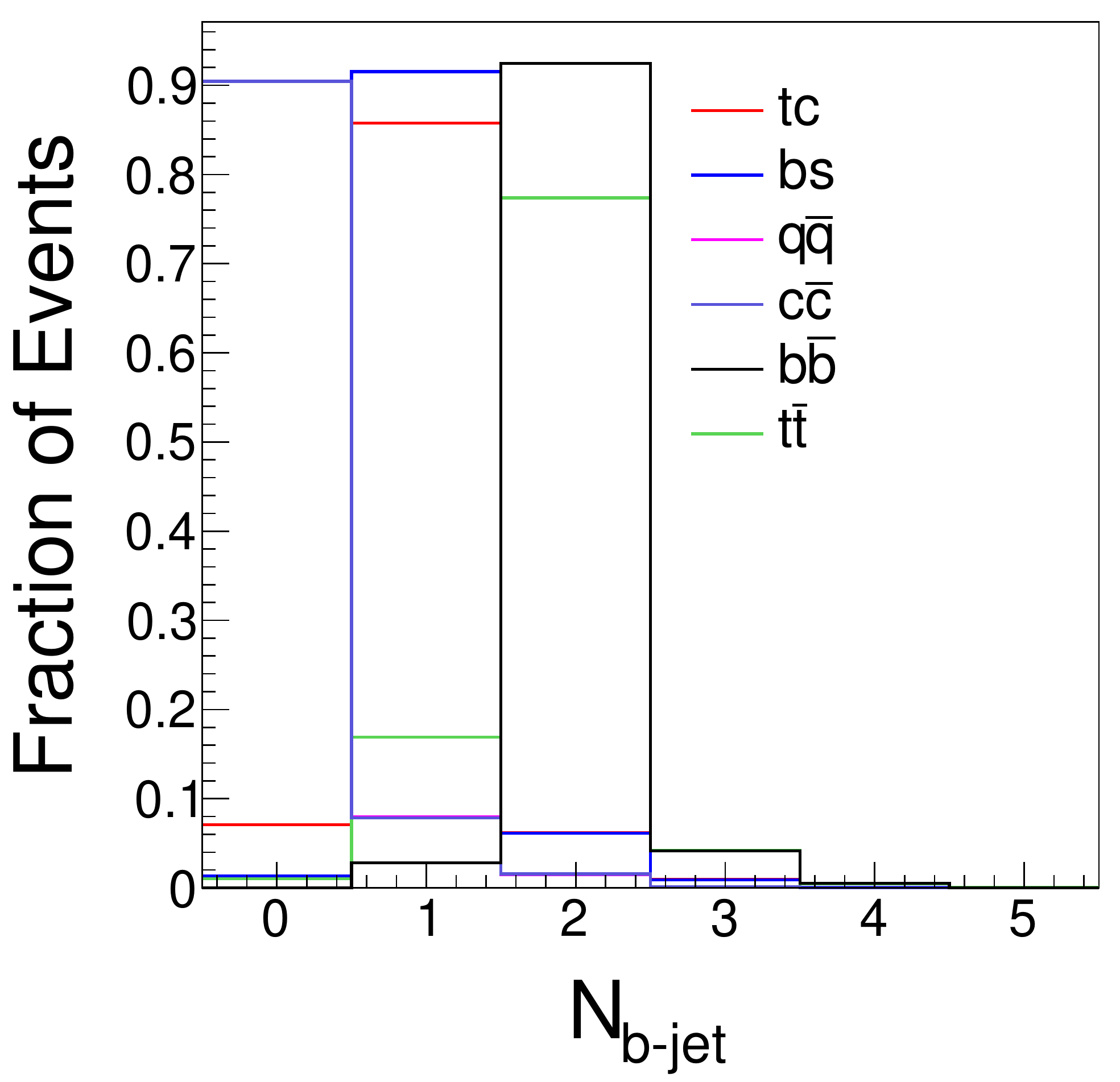}
     \caption{}
     \label{fig:nbjets}
  \end{subfigure}
  \begin{subfigure}[t]{0.45\textwidth}
     \includegraphics[width=\linewidth]{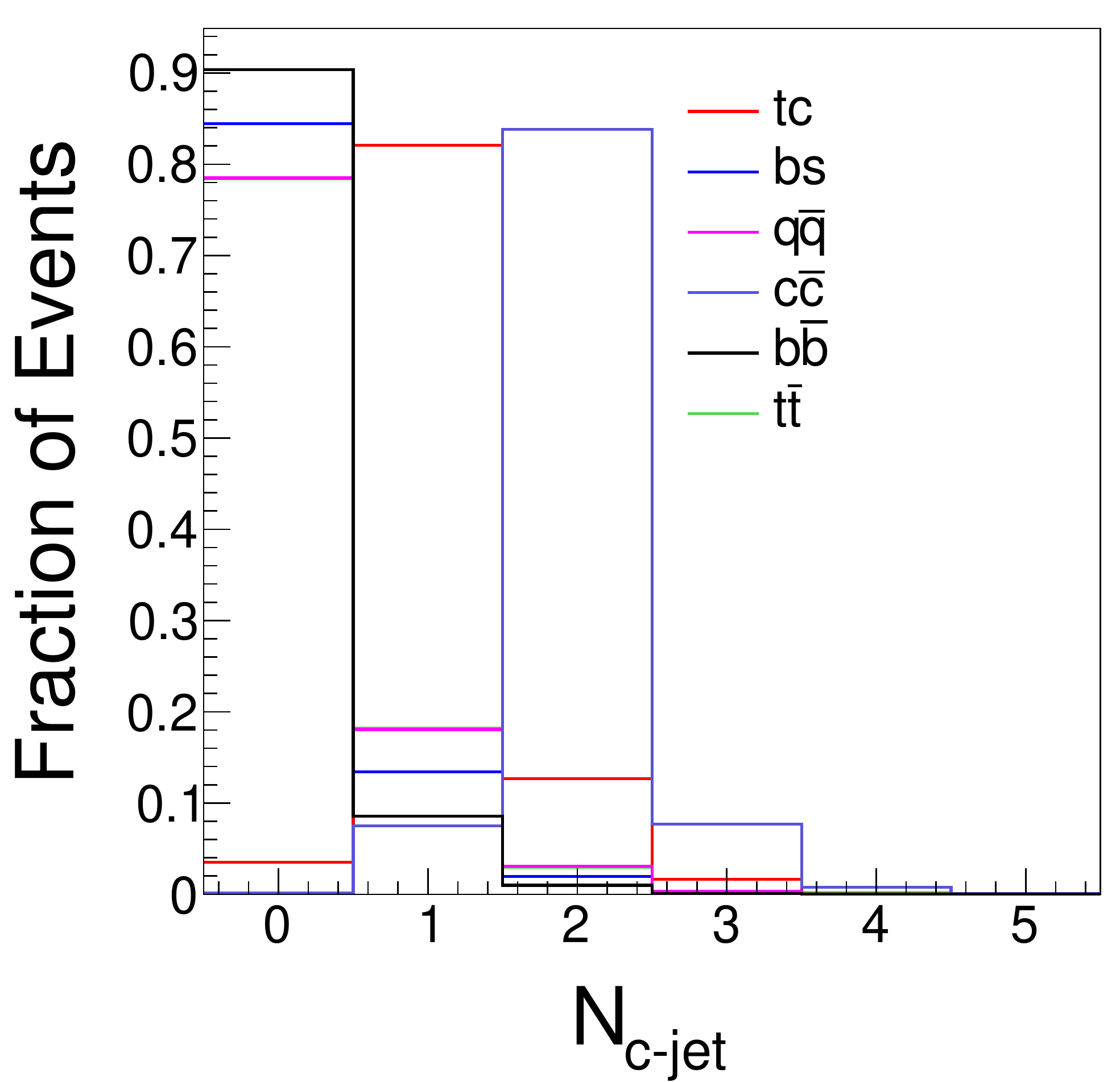}
     \caption{}
     \label{fig:ncjets}
  \end{subfigure}
  \begin{subfigure}[t]{0.45\textwidth}
     \includegraphics[width=\linewidth]{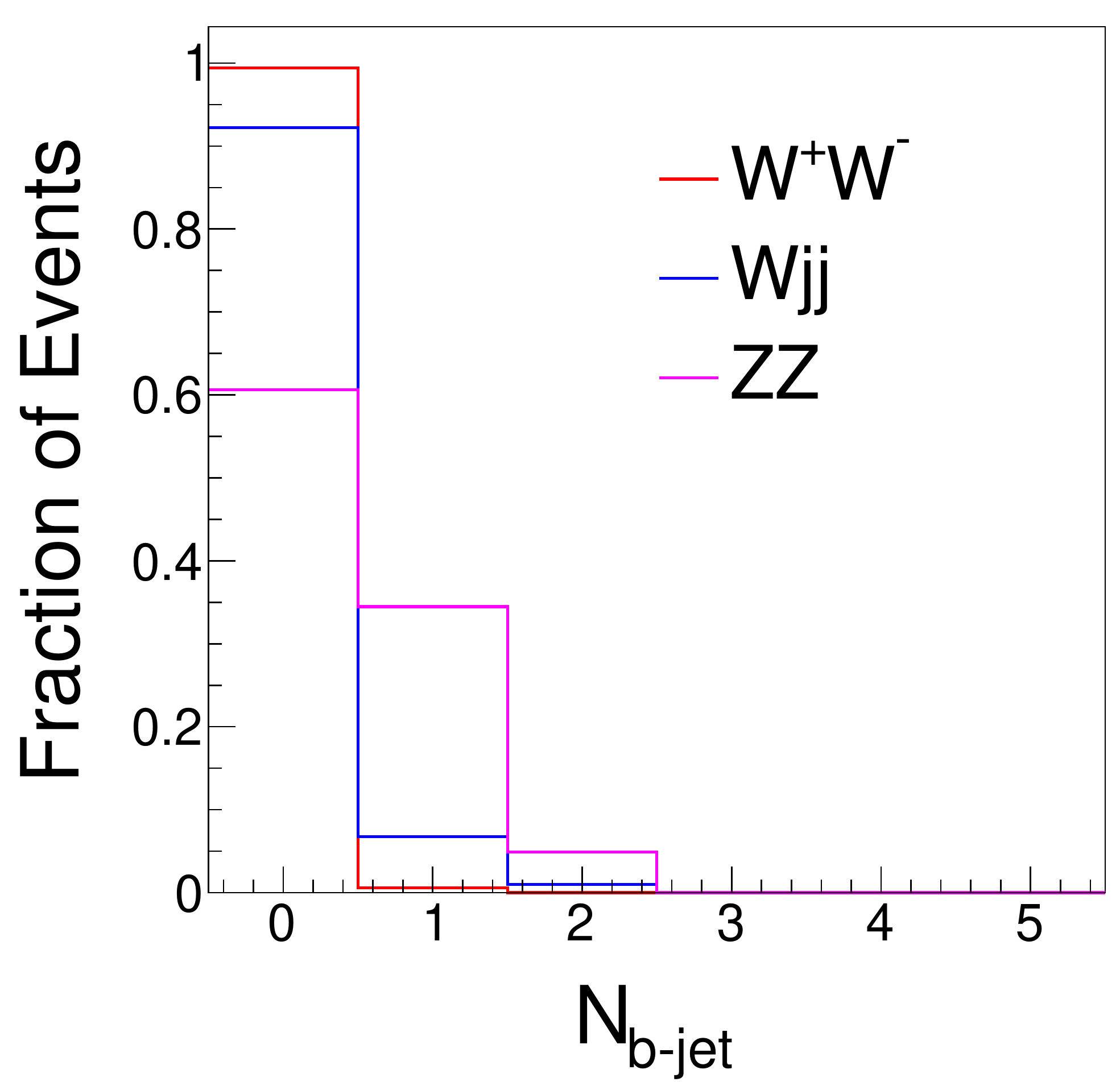}
     \caption{}
     \label{fig:nbjets_wz}
  \end{subfigure}
  \begin{subfigure}[t]{0.45\textwidth}
     \includegraphics[width=\linewidth]{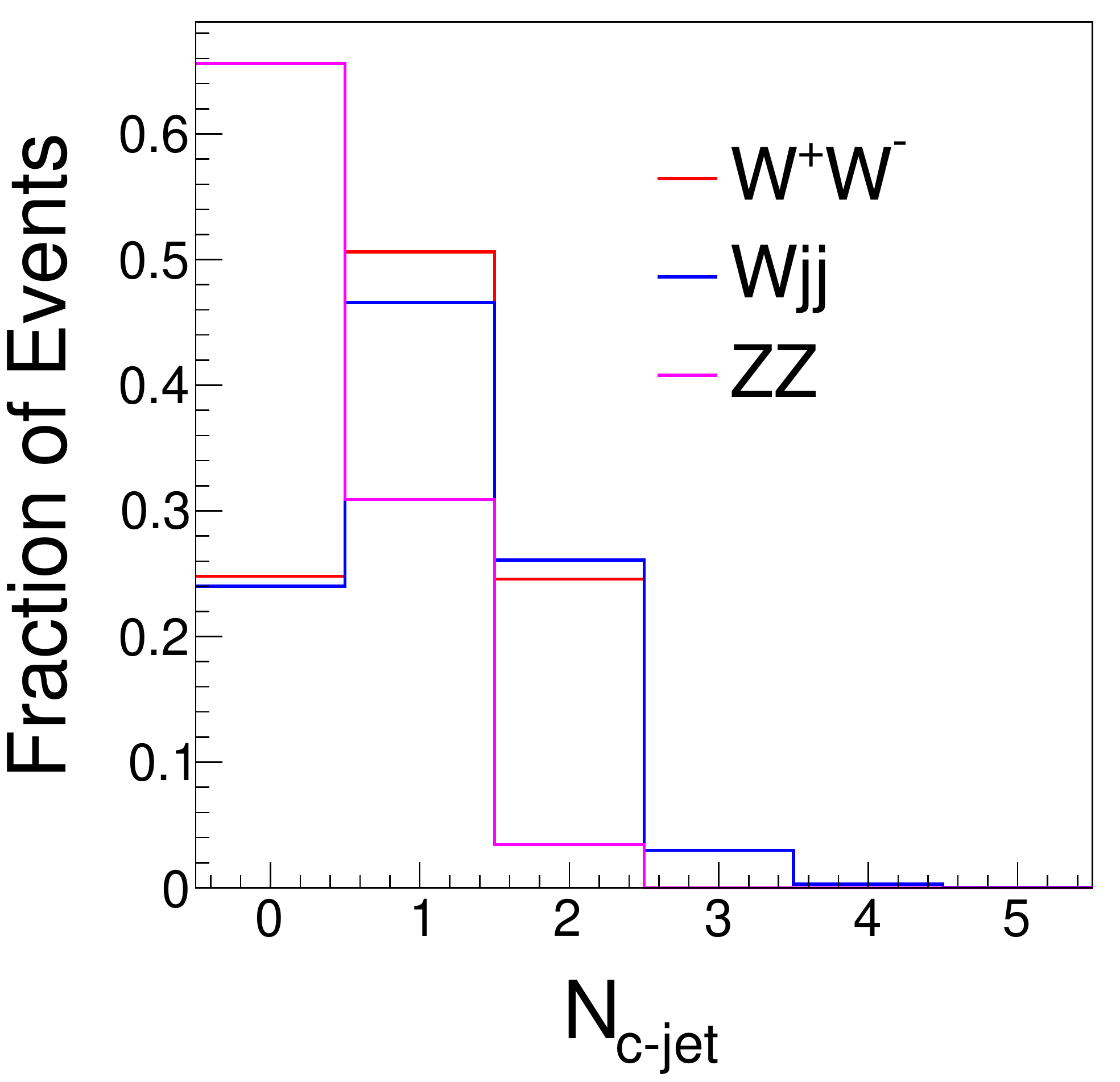}
     \caption{}
     \label{fig:ncjets_wz}
  \end{subfigure}
  \caption{The number of (a) true $b$-jets and (b) true $c$-jets in signals and backgrounds with quarks are displayed. As a reference, the number of (c) true $b$-jet and (d) true $c$-jet in the background with gauge boson are displayed.}\label{fig:bcjet}
\end{figure}

Fig.~\ref{fig:nbjets} and Fig.~\ref{fig:ncjets} show the number of true $b$-jets and $c$-jets for signals and backgrounds with quarks.
Obviously, a $b$-jet and a $c$-jet are found in the $t\bar{c}(\bar{t}c)$ signal. 
For backgrounds with two b quarks ($t\bar{t}$ and $b\bar{b}$), two $b$-jets can be found in most events.
For the $c\bar{c}$ backgrounds, most events include two $c$-jets.
Most of the $q\bar{q}$ events do not have both $b$- and $c$-jet, 
but a small fraction of such events can have $b$- or $c$-jets in the final states.  
For example, in the parton shower, a quark has a certain probability to radiate a gluon, 
and subsequently, this gluon may split to heavy quarks like $b\bar{b}$ and $c\bar{c}$. 
Such a kind of process is easier to occur for an energetic light quark,  
which may lead to an increase in the mistagging rate of light quarks. 
For backgrounds with gauge bosons, as we plot in Fig.~\ref{fig:nbjets_wz} and Fig.~\ref{fig:ncjets_wz}, 
$ZZ$, $W^+W^-$ or $W^{\pm}jj$ can decay to $b$ or $c$ quark, 
so we also observe some $b$- or $c$-jets in these events.

With this flavor tagging information, we can implement a $b$-tagging cut. 
In our analysis, the $b$-tagging rate is $\epsilon_{b}=0.7$, 
while the mistagging rate is $\epsilon_{c}=0.1$ and $\epsilon_{q}=0.01$ for $c$-jets and light jets, respectively.

Below we introduce some simple cuts to separate signal and background events:
\begin{itemize}
  \item Cut1: $N_{jet}=2$, $N_{lepton}=0$ and $M_{jj}>8$ TeV,
  \item Cut2: $150$ GeV$<M(HJ)<200$ GeV,
  \item Cut3: $M(LJ)<75$ GeV,
  \item Cut4: the heavy jet is b-tagged and the light jet is not tagged.
\end{itemize}
The results of the cut flows are listed in Table.~\ref{table:tc:cut}. 
As we can see, the cuts for the invariant masses of heavier and lighter jets are efficient to reduce backgrounds with heavy particles.
The $b$-tagging cut is efficient to reduce backgrounds with $W$ boson. 
After these cuts, the huge backgrounds are reduced successfully, 
and the signal events can be observed with remarkable significance. 

From Table \ref{table:tc:cut}, it is interesting to note that the $ZZ$ background is zero after Cut2, 
but few $WW$ events survive.
It is because Pythia8 has different treatments to the $ZZ\to q_1\bar{q}_1q_2\bar{q}_2$ and $W^+W^-\to q_1\bar{q}_2q_3\bar{q}_4$. 
For the $ZZ$ events, there is no confusion that $q_1\bar{q}_1$ and $q_2\bar{q}_2$ are combined to a color singlet, respectively. 
For the $W^+W^-$ events, color reconnection can lead to the formation of alternative ``flipped'' singlets $q_1\bar{q}_4$ and $q_3\bar{q}_2$. 
In this case, few $W^\pm$ propagators are forced off the mass shell~\cite{Sjostrand:1993hi}. 
In this work, we use the default tune of Pythia8, which is called Monash 2013~\cite{Skands:2014pea}. 
The default model for color reconnection is so-called MPI-based model~\cite{Sjostrand:1987su}. 
Pythia8 also provides models intended for $e^+e^-\to W^+W^-$~\cite{Sjostrand:1993hi,Sjostrand:1993rb}. 
The choice of color reconnection model has some effects to our analysis, 
but do not change our final conclusion. 

\begin{center}
\begin{table}
  \begin{center}
  \begin{tabular}{|c|c|c|c|c|c|c|c|}
  \hline
  &  No Cuts  &     Cut1   &  Cut2  & Cut3 & Cut4 &  $S/B$  &    $\sigma$   \\
  \hline
  $\mu^+\mu^-\to tc$   &  $847$   & $389$  &  $247$  &  $166$ & $101$  &  \multirow{8}{*}{0.38} & \multirow{8}{*}{5.26} \\
  $\mu^+\mu^-\to t\bar{t}$     &   $7.15\times{10^3}$  &       $3323$  &  $2534$ & $90$ & $35$  & & \\
  $\mu^+\mu^-\to q\bar{q}$     &   $3.56\times{10^4}$   & $1.65\times{10^4}$  &    $1696$    &  $1167$ & $75$ &   &  \\
  $\mu^+\mu^-\to c\bar{c}$     &   $1.73\times{10^4}$  &  $7663$  &    $788$    &  $559$ & $70$ &   &  \\
  $\mu^+\mu^-\to b\bar{b}$     &   $9137$  & $3744$  &    $386$    &  $276$ & $53$ &   &  \\
  $\mu^+\mu^-\to W^+W^-$  & $2.51\times{10^5}$   & $2.39\times{10^5}$  & $1181$ &  $186$ & $0$ & & \\
  $\mu^+\mu^-\to Wjj$  & $3.55\times{10^5}$  & $1.11\times{10^4}$  &  $976$   &  $603$ & $35$  & & \\
  $\mu^+\mu^-\to ZZ$  & $1.52\times{10^4}$   & $1.50\times{10^4}$  &  $0$   &  $0$ & $0$  &  & \\
  \hline
  \end{tabular}
  \end{center}
  \caption{The number of events before and after each cut are listed, where $\sigma$ is defined as $\sigma=S/\sqrt{S+B}$. We use BP3 and Model I as theoretical inputs for the signal. We assume the luminosity is $10$ ab$^{-1}$ at 10 TeV muon collider. The details of these cuts are described in the text.\label{table:tc:cut}}
\end{table}
\end{center}

For a purpose of comparison and contrast, we also perform an analysis of the process $\mu^+\mu^-\to b\bar{s}(\bar{b}s)$. We introduce the following cuts to separate signal and background events 
\begin{itemize}
  \item Cut1: $N_{jet}=2$, $N_{lepton}=0$, and $M_{jj}>8$ TeV,
  \item Cut2: $M(HJ)<75$ GeV,
  \item Cut3: One jet is b-tagged.
\end{itemize} 
The results of cut flows are listed in Table.~\ref{table:bs:cut}.

As we discussed above, there should be no massive jets in such signal events, 
so we can veto the heavier jets by demanding the heavier jet should not have too much larger jet mass.
The jet mass cut of the heavier jets can work to reject backgrounds like $t\bar{t}$, $WW$, and $ZZ$.
The remained backgrounds with $W$ boson can be further reduced by applying a b-tagging cut. 
After these cuts, it is observed that heavy flavor final states will be the dominant background which leads to a small $S/B$.

\begin{center}
\begin{table}
  \begin{center}
  \begin{tabular}{|c|c|c|c|c|c|c|}
  \hline
  &  No Cuts  &     Cut1 & Cut2 & Cut3 & $S/B$ &  $\sigma$  \\
  \hline
  $\mu^+\mu^-\to bs$     &   $1533$  &   $693$  &  $309$ & $214$   &   \multirow{8}{*}{0.08}   &  \multirow{8}{*}{3.99}  \\
  $\mu^+\mu^-\to t\bar{t}$     &   $7152$    & $3323$   &  $1$  & $1$  & & \\
  $\mu^+\mu^-\to q\bar{q}$     &   $3.56\times{10^4}$    &   $1.65\times{10^4}$ & $7128$    &  $356$ &   &  \\
  $\mu^+\mu^-\to c\bar{c}$     &   $1.73\times{10^4}$   &    $7663$ & $3368$    &  $735$  &   &  \\
  $\mu^+\mu^-\to b\bar{b}$     &   $9137$    &  $3744$ &   $1662$    &  $673$  &   &  \\
  $\mu^+\mu^-\to W^+W^-$  & $2.51\times{10^5}$     &  $2.39\times{10^5}$ &  $4151$ &  $462$ & & \\
  $\mu^+\mu^-\to Wjj$  & $3.55\times{10^5}$   & $1.11\times{10^4}$  &  $3957$   &  $373$  & & \\
  $\mu^+\mu^-\to ZZ$  & $1.52\times{10^4}$  & $1.50\times{10^4}$ &  $149$   &  $63$  &  & \\
  \hline
  \end{tabular}
  \end{center}
  \caption{The number of events before and after each cut is listed. We use BP3 and Model I as theoretical inputs for the signal. We assume the luminosity is $10$ ab$^{-1}$ at 10 TeV muon collider. The details of these cuts are described in the text.\label{table:bs:cut}}
\end{table}
\end{center}

\begin{figure}[htbp]
  \centering
  \captionsetup[sub]{font=large}
  \begin{subfigure}[t]{0.45\textwidth}
     \includegraphics[width=\linewidth]{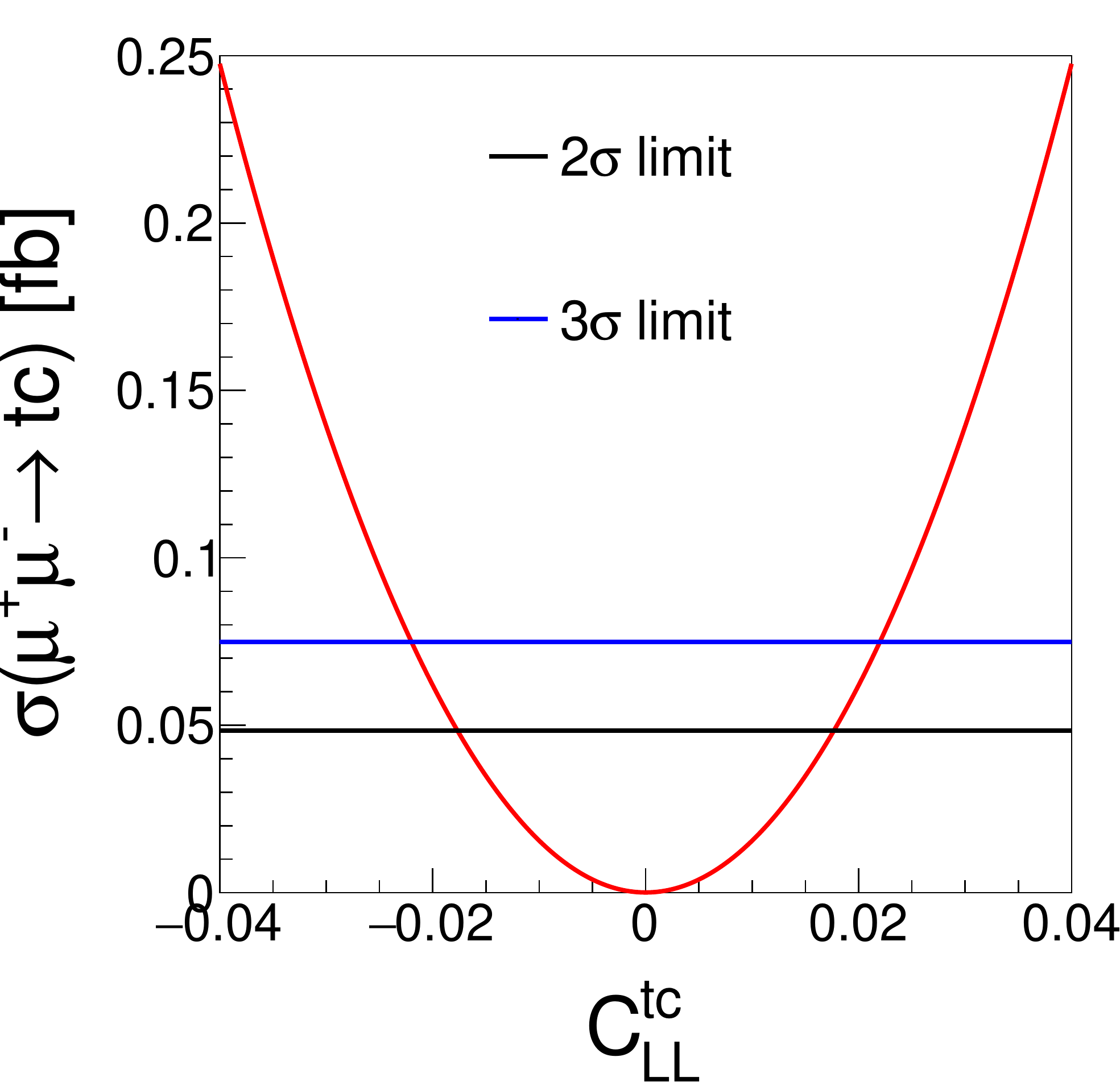}
     \caption{}
     \label{fig:clltc1d}
  \end{subfigure}
  \begin{subfigure}[t]{0.45\textwidth}
     \includegraphics[width=\linewidth]{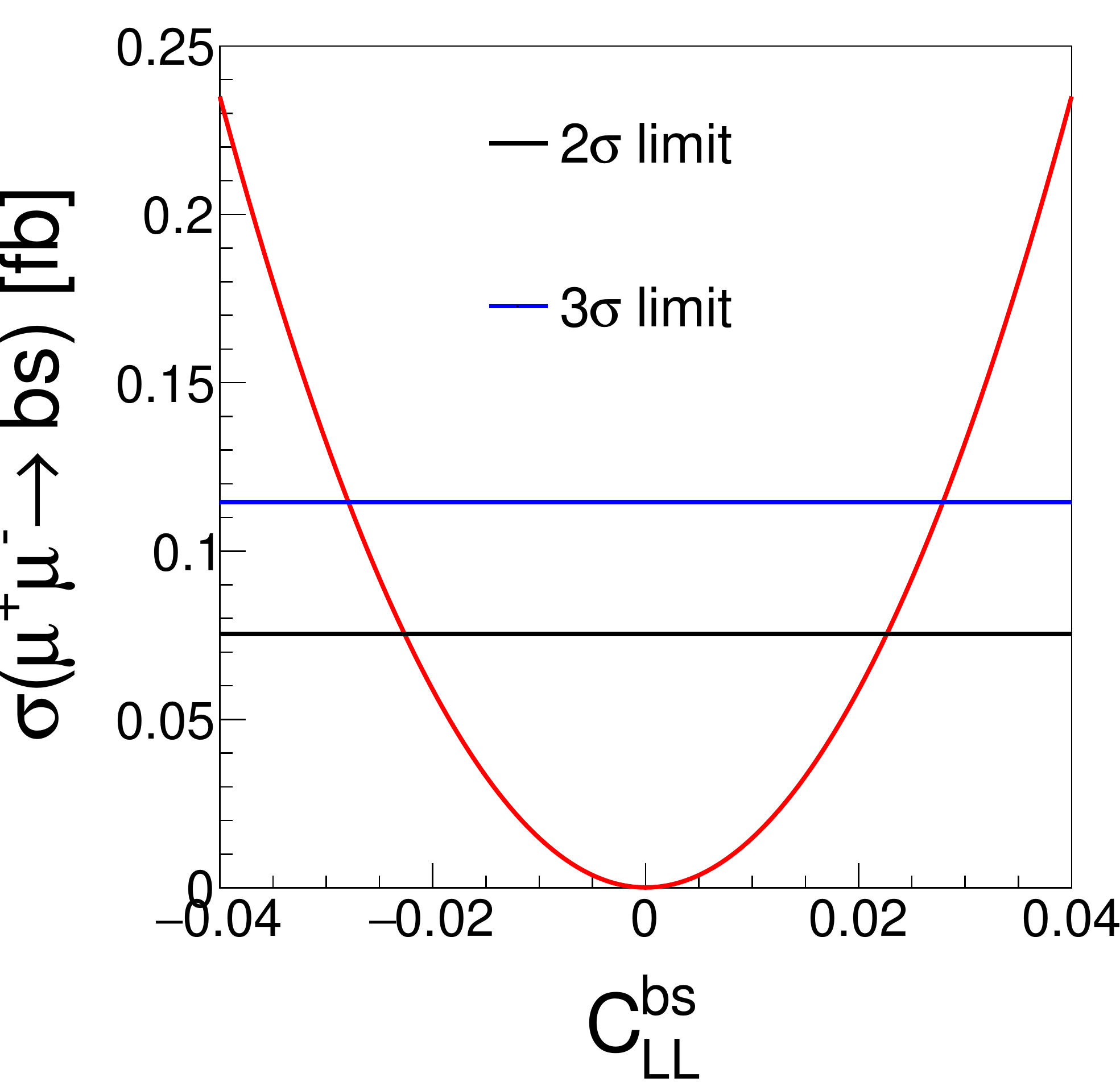}
     \caption{}
     \label{fig:cllbs1d}
  \end{subfigure}
  \caption{The $2\sigma$ and $3\sigma$ bounds of (a) $C^{tc}_{LL}$ and (b) $C^{bs}_{LL}$ at 10 TeV muon collider with luminosity $\mathcal{L}=10$ ab$^{-1}$ are shown.}\label{fig:cll1d}
\end{figure}

In Fig.~\ref{fig:cll1d}, we demonstrate the $2\sigma$ and $3\sigma$ bounds on $C^{X}_{LL}$ from the measurement of total cross sections. Based on our analysis, $C^{tc}_{LL}$ can be constrained to $[-1.8\times 10^{-2}, +1.8\times 10^{-2}]$ ($[-2.2\times 10^{-2}, +2.2\times10^{-2}]$) at $2\sigma$ ($3\sigma$) limit at 10 TeV muon collider with luminosity $\mathcal{L}=10$ ab$^{-1}$, while 
$C^{bs}_{LL}$ can be constrained to $[-2.3\times 10^{-2}, +2.3 \times 10^{-2}]$ ($[-2.8\times 10^{-2}, +2.8\times 10^{-2}]$) at $2\sigma$ ($3\sigma$) limit. 

\begin{figure}[htbp]
  \centering
  \captionsetup[sub]{font=large}
  \begin{subfigure}[t]{0.45\textwidth}
     \includegraphics[width=\linewidth]{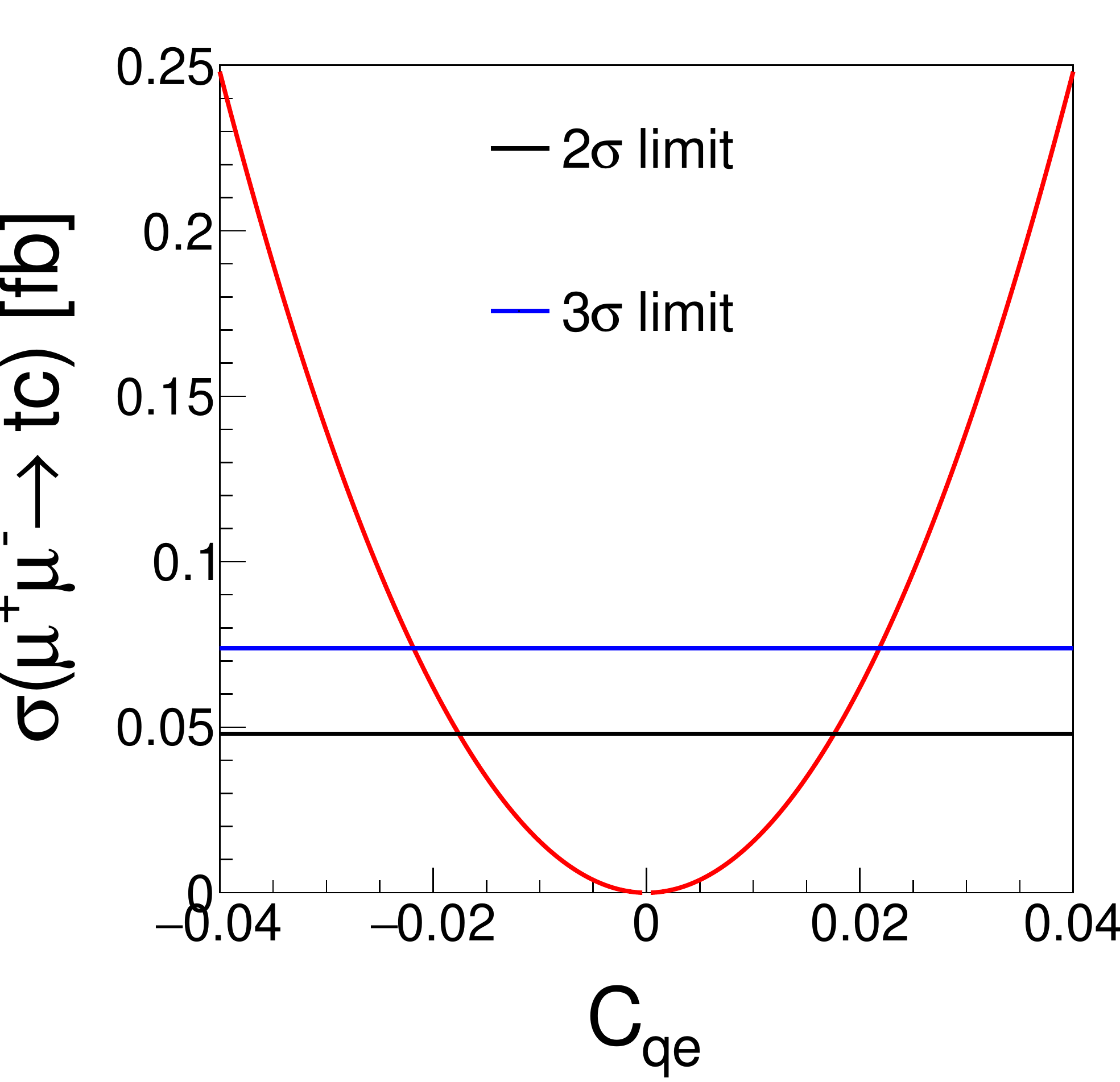}
     \caption{}
     \label{fig:clltc1d}
  \end{subfigure}
  \begin{subfigure}[t]{0.45\textwidth}
     \includegraphics[width=\linewidth]{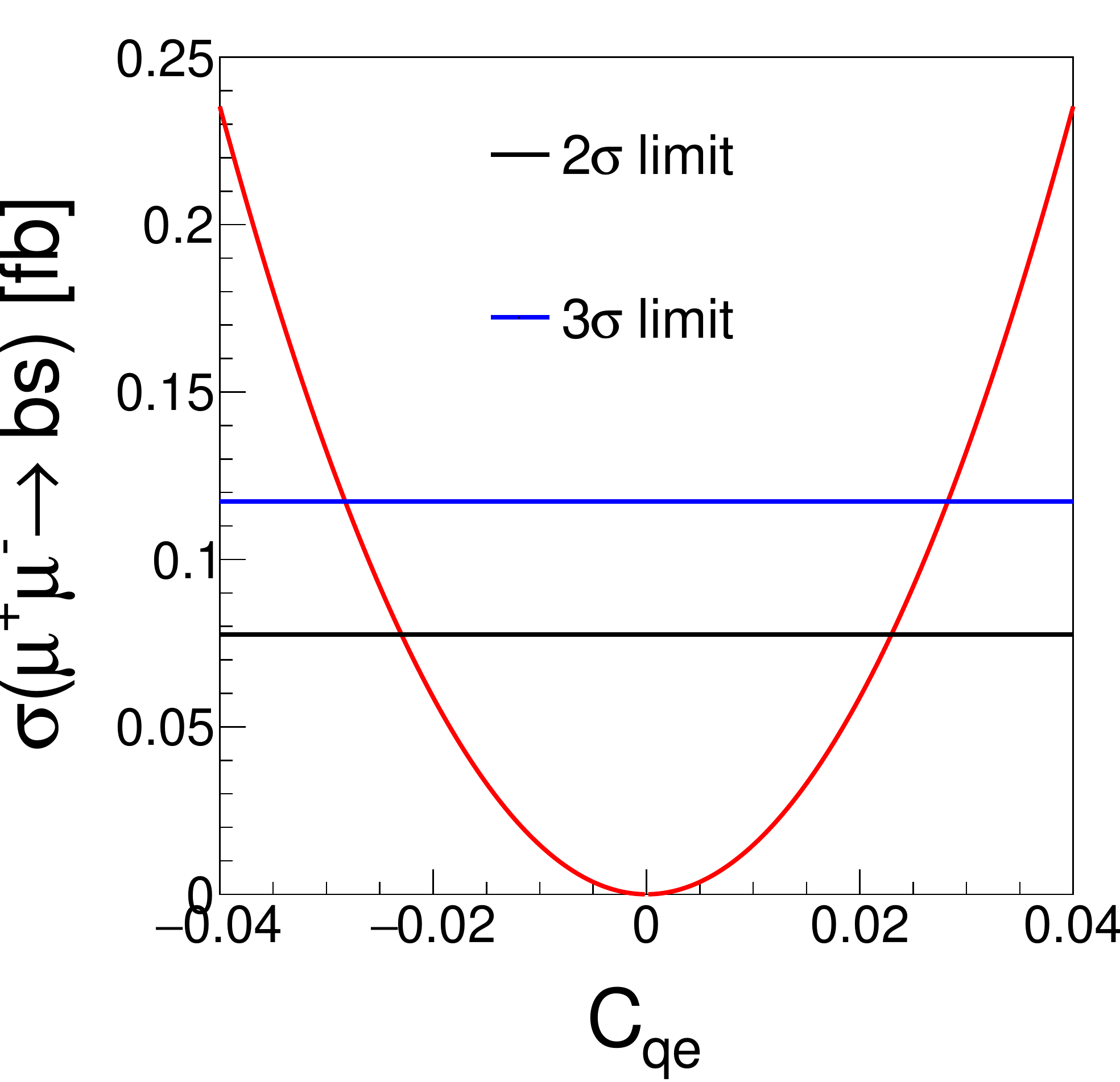}
     \caption{}
     \label{fig:cllbs1d}
  \end{subfigure}
  \caption{The $2\sigma$ and $3\sigma$ bounds of $C_{qe}$ obtained by measuring (a) $tc$ and (b) $bs$ final state at 10 TeV muon collider with luminosity $\mathcal{L}=10$ ab$^{-1}$ are shown.}\label{fig:cqe1d}
\end{figure}

Similarly, we implement the same cuts to analyze the case when only $C_{qe}$ is switched on. 
In Fig.~\ref{fig:cqe1d}, we show the $2\sigma$ and $3\sigma$ bounds on $C_{qe}$. 
There is no wonder that these bounds are close to that we obtain for $C^{X}_{LL}$ when only the information of the total cross section is utilized.

\begin{figure}
    \includegraphics[width=0.45\textwidth]{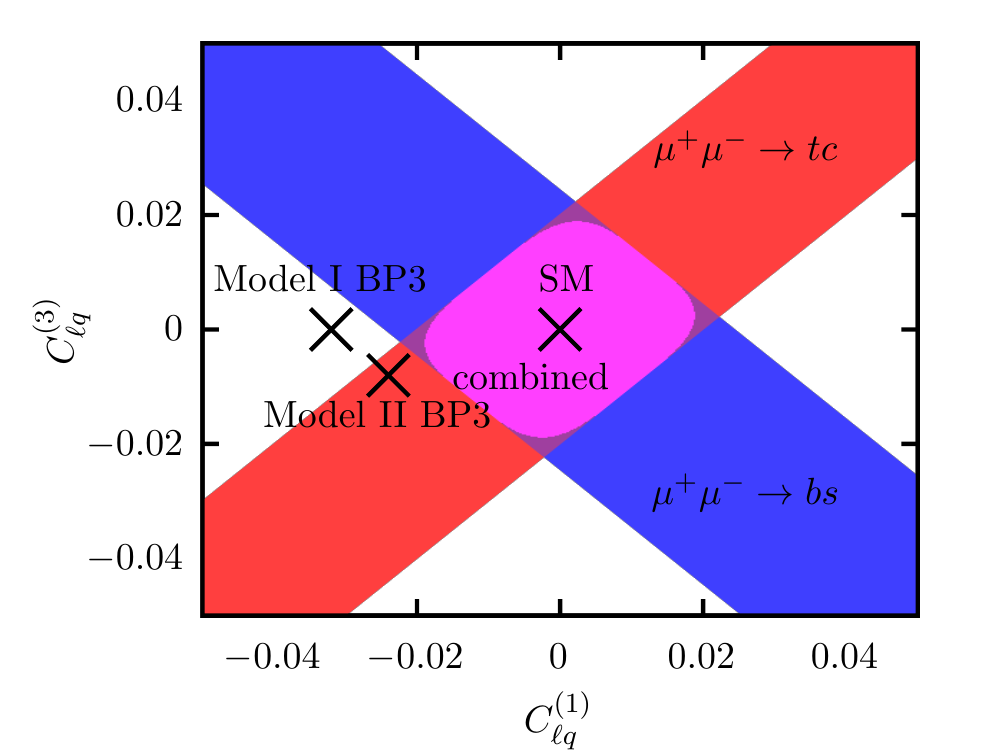}
    \includegraphics[width=0.45\textwidth]{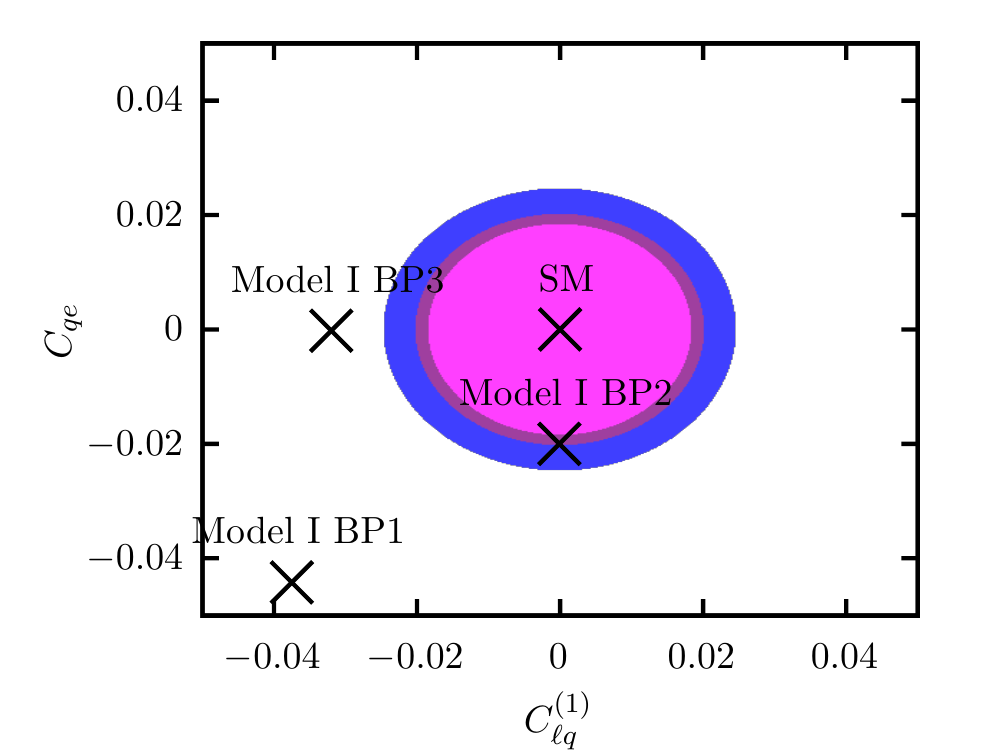}
    \includegraphics[width=0.45\textwidth]{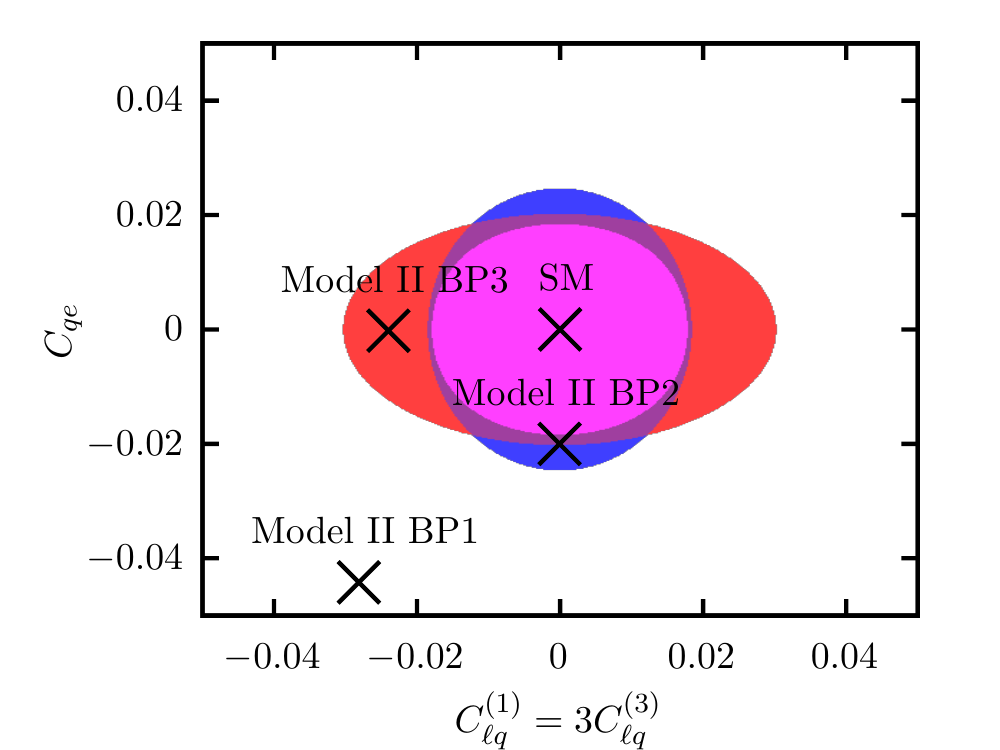}
\caption{The 2-D 95\% CL bounds are shown for the case of $C_{qe}=0$(top left), $C_{\ell q}^{(3)}=0$(top right), $C_{\ell q}^{(1)}=3C_{\ell q}^{(3)}$(bottom).}\label{fig:cll2d}
\end{figure}

In Fig. \ref{fig:cll2d}, we show the 2D constraints from $\mu^+\mu^-\to tc$(blue area), $\mu^+\mu^-\to bs$(red area), and their combination(magenta area).
In the top left panel of Fig. \ref{fig:cll2d}, we show the constraints under the assumption that $C_{qe}=0$.
We note that only BP3 satisfies this assumption in our benchmark points,
while the underlying models are unknown.
Constraints from $\mu^+\mu^-\to bs$ can probe whether the four-fermion operators are in charge of the beyond the standard model rare B decay processes,
while $\mu^+\mu^-\to tc$ are essential to distinguish different underlying models.
In the top right panel of Fig. \ref{fig:cll2d},
we show the constraints under the assumption that $C_{lq}^{(3)}=0$, i.e. in Model I.
We can see that in this case both $\mu^+\mu^-\to tc$ and $\mu^+\mu^-\to bs$ have similar dependence on $C_{qe}$ and $C_{lq}^{(1)}$, though the former provides stronger constraints.
Combining both channels we can distinguish BP1, BP2, and BP3.
In the bottom panel of Fig. \ref{fig:cll2d},
we show the constraints under the assumption that $C_{lq}^{(1)}=3C_{lq}^{(3)}$, i.e. Model II.
We can see that in this case $\mu^+\mu^-\to tc$ is more sensitive to $C_{qe}$, while $\mu^+\mu^-\to bs$ is more sensitive to $C_{lq}^{(1)}$.
Both channels are complementary and combining them we can distinguish different benchmark points.

As we have explained in section 2, 
the constraints on $C^{(1)}_{lq}+C^{(3)}_{lq}$ and $C_{qe}$ can also be applied to $C_{ld}$ and $C_{ed}$.  
It is difficult to distinguish the left-handed and right-handed contributions with unpolarized muon beams.
If the B decays data are in favor of the operators $O^{\prime}_{9}$ and $O^{\prime}_{10}$, 
correspondingly we should observe deviations in $\mu^+\mu^-\to bs$ process, 
but not in $\mu^+\mu^-\to tc$ process. 
To further distinguish the contributions of left-handed or right-handed, the measurement of cross sections with polarized muon beams might be necessary. 

\section{Summary and Discussion}\label{Sec:conc}
For the signal processes $\mu^+ \mu^- \to b s$ \cite{Altmannshofer:2022xri}, it has been revealed that b tagging is crucial to reject background events from $jj$ final states. The dominant background events are $j j$. In order to extract the information on the Wilson coefficients $C_{9}$ and $C_{10}$, it is found that the polarized muon beams and the measurement of forward-backward asymmetry of the final state are needed. 

In this work, instead of assuming polarized muon beams and charge tagging of a final state, we propose to measure the signal processes $\mu^+ \mu^- \to t c$. Then it is found that the major task is to reject $t \bar{t}$ and $ W W$ events, which might be easier to pick out signal events. It is also found that the weak radiation $W j j $ is large and might be a relevant background.

It is also found that in order to resolve the signal events, detectors with high granularity are needed since top quarks and W bosons in the hadronic final states are around 5 TeV and they are highly boosted objects. 
In order to capture the substructure of these massive jets, 
the cone parameter should be set as around  $0.09-0.1$ when the collision energy is assumed to be $\sqrt{s}=10$ TeV, 
which is much smaller compared with the cone parameter adopted as $R=0.4$ or $0.5$ at the LHC. 
To extract signal events from large background events, 
a refined analysis of the TeV region jet substructure~\cite{Butterworth:2008iy,Kaplan:2008ie,Kilian:2021whd} should be applied to achieve much better performance. 
It is expected that a modern top tagger technique can improve the top jet identification and reject the W/Z jets.

To distinguish new physics models, it is found that the measurement of  $\mu^+ \mu^- \to tc$ can pinpoint Model I and Model II. Model III can be discovered due to the resonance near a few TeV regions. Model IV could be favored if there is no signal of the process $\mu^+ \mu^- \to t c$ are measured.

Finally, we have some comments on the systematic uncertainties. In real data analysis, the systematic uncertainties should include uncertainties of collision energy, uncertainties of luminosity of incoming muon beams, statistical uncertainty, theoretical uncertainties, etc. In this work, the statistical uncertainty is assumed to be dominant. 
There are other extra theoretical sources which might also contribute to the systematic uncertainties: 
\begin{itemize}
  \item The first one is the theoretical uncertainty from higher order corrections. 
In particular, the electroweak corrections will become important at a multi-TeV muon collider, 
so we have to consider background like $Wjj$.
The automation with full electroweak corrections is still challenged and under development. 
Our tree-level results are preliminary.
  \item The second one is the uncertainty from parton shower and hadronization. 
As we have seen, the color reconnection effect lead to some interesting behaviour of $W^+W^-$ background. 
The description of such kind of effects is strongly depending on the tuning of MC generators.
  \item In our study, the initial state radiation (ISR) is switched off. 
One can expect that the ISR of multi-TeV muon beam will be important, 
and the radiation of massive particles (W, Z and Higgs) should be considered. 
However, our MC tools are incomplete to give an exact description. 
\end{itemize}
To explore these theoretical uncertainties are also the challenges for the future muon collider. 
We still have many works to do.

	\begin{acknowledgments}
                Z.J. Zhao has been partially supported by a 
                China and Germany Postdoctoral Exchange Program between the Office of China Postdoctoral Council (OCPC) and DESY,
		, and partially supported by the Natural Science
		Foundation of China under the grant No. 11875260. 
		S.C. Sun is supported by the National Natural Science Foundation of China, No.12105013.
		Q.S. Yan is supported by the Natural Science Foundation of China
		under the grant No.  11475180 and No. 11875260.
		X.R. Zhao is supported by the Italian Ministry of Research (MUR) under grand PRIN 20172LNEEZ.

	\end{acknowledgments}

\appendix

\section{SMEFT Renormalization Group Equation}\label{smeftrge}
The RGE of SMEFT Wilson coefficients can be written as 
\begin{eqnarray}
  \frac{dC_i}{d\ln{\mu}} &=& \frac{1}{16\pi^2}\beta_i. \label{beta:definition}
\end{eqnarray}
All 1-loop $\beta$ functions of operators in Warsaw basis have been derived in Ref.~\cite{Celis:2017hod}. 

The $\beta$ functions of operators~\ref{O1lq}$\sim$\ref{Oed} are
\footnote{We only consider the terms are non-zero when $(p,r,s,t)=(2,2,2,3)$ ($(2,3,2,2)$ for $C_{qe}$). This is the same for the LEFT case.}
\begin{eqnarray}
   \left[\beta^{(1)}_{lq}\right]_{prst} &=& \frac{2}{3}{g^\prime}^2\left([C^{(1)}_{lq}]_{wwst}+[C_{qe}]_{stww}\right)\delta_{pr}-{g^\prime}^2[C^{(1)}_{lq}]_{prst}  \nonumber \\
   &&+9g^2[C^{(3)}_{lq}]_{prst}+\frac{1}{2}[\Gamma_u^\dagger\Gamma_u]_{vt}[C^{(1)}_{lq}]_{prsv}, \label{beta:lq1}  \\
   \left[\beta^{(3)}_{lq}\right]_{prst} &=& \frac{2}{3}g^2[C^{(3)}_{lq}]_{wwst}\delta_{pr}+3g^2[C^{(1)}_{lq}]_{prst}-(6g^2+{g^\prime}^2)[C^{(3)}_{lq}]_{prst}  \nonumber \\ 
   && +\frac{1}{2}[\Gamma_u^\dagger\Gamma_u]_{vt}[C^{(3)}_{lq}]_{prsv}, \label{beta:lq3} \\
   \left[\beta_{qe}\right]_{prst} &=& \frac{4}{3}{g^{\prime}}^2\left([C^{(1)}_{lq}]_{wwpr}+[C_{qe}]_{prww}\right)\delta_{st}+2{g^{\prime}}^2[C_{qe}]_{prst} \nonumber  \\
   && +\frac{1}{2}[\Gamma_u^\dagger\Gamma_u]_{vr}[C^{(1)}_{lq}]_{pvst}, \label{beta:qe} \\
   \left[\beta_{ld}\right]_{prst}  &=& \frac{2}{3}{g^\prime}^2[C_{ld}]_{wwst}\delta_{pr}+\frac{2}{3}{g^\prime}^2[C_{ed}]_{wwst}\delta_{pr}-2{g^\prime}^2[C_{ld}]_{prst} \\
  \left[\beta_{ed}\right]_{prst} &=& -\frac{2}{3}{g^\prime}^2\left(2[-C_{ld}-C_{ed}]_{wwst}\right)\delta_{pr}+4{g^{\prime}}^2[C_{ed}]_{prst}
\end{eqnarray}
where $g$ and $g^\prime$ are the gauge coupling of $SU(2)$ and $U(1)$, respectively. 
$\Gamma_{u}$ is the $3\times 3$ Yukawa mass matrix of u-type quarks. 
For simplicity, we only consider the top quark is massive, 
and only the element $\Gamma_u(3,3)=1$ is non-zero.
Note the Eq.~\ref{beta:lq1}$\sim$\ref{beta:qe} only contains the most important contributions and mixing of the corresponding operators. 
The mixing of the full set of operators is beyond the scope of this work. 

The 1-loop RGE running of SM parameters is given by these $\beta$ functions:
\begin{eqnarray}
  \beta_{g} &=& -\frac{19}{6}g^3,  \\
  \beta_{g^\prime} &=& \frac{41}{6}{g^\prime}^3,  \\
  \beta_{g_s} &=& -7g^2_s, \\
  \left[\beta_{\Gamma_u}\right]_{33} &=& \frac{9}{4}g^2\Gamma_u(3,3)-\frac{17}{12} {g^\prime}^2\Gamma_u(3,3)-8g^2_s\Gamma_u(3,3)+\frac{9}{2}\Gamma^3_u(3,3).
\end{eqnarray}

Our simplified RGE running has been compared with two tools: DSixTools~\cite{Celis:2017hod,Fuentes-Martin:2020zaz} and Wilson~\cite{Aebischer:2018bkb}.
The differences between our results and these tools are below $1\%$.

\section{LEFT Renormalization Group Equation}\label{leftrge}

The definition of RGE of LEFT is the same as Eq.~\ref{beta:definition}, but $C_i$ are replaced by $L_i$.
The $\beta$ functions of operator~\ref{QVLLed}$\sim$\ref{QVRRed} are
\begin{eqnarray}
  \left[\beta^{V,LL}_{ed}\right]_{prst} &=& \frac{4}{3}e^2q^2_e\delta_{pr}\left([L^{V,LL}_{ed}]_{wwst}+[L^{V,LR}_{de}]_{stww}\right)+12e^2q_eq_d[L^{V,LL}_{ed}]_{prst},  \label{beta:VLLed} \\
  \left[\beta^{V,LR}_{de}\right]_{prst} &=& \frac{4}{3}e^2q^2_e\delta_{st}\left([L^{V,LL}_{ed}]_{wwpr}+[L^{V,LR}_{de}]_{prww}\right)-12e^2q_eq_d[L^{V,LR}_{de}]_{prst}, \label{beta:VLRde} \\
  \left[\beta^{V,LR}_{ed}\right]_{prst} &=& \frac{4}{3}e^2q^2_e\delta_{pr}\left([L^{V,LR}_{ed}]_{wwst}+[L^{V,RR}_{ed}]_{wwst}\right)-12e^2q_eq_d[L^{V,LR}_{ed}]_{prst}, \label{beta:VLRed} \\
  \left[\beta^{V,RR}_{ed}\right]_{prst} &=& \frac{4}{3}e^2q^2_e\delta_{pr}\left([L^{V,LR}_{ed}]_{wwst}+[L^{V,RR}_{ed}]_{wwst}\right)+12e^2q_eq_d[L^{V,RR}_{ed}]_{prst}, \label{beta:VRRed} 
\end{eqnarray}
where $e$ is the coupling constant of QED. 
$q_e=-1$ and $q_d=-1/3$ are the charges of lepton and d-type quark, respectively. 

At the low energy scale, the 1-loop running of QCD and QED coupling is given by the following $\beta$ functions:
\begin{eqnarray}
  \beta_{g_s} &=& -\frac{23}{3}g^3_s,  \\
  \beta_{e} &=& \frac{80}{9}e^3, 
\end{eqnarray}

		\bibliographystyle{JHEP}
		\bibliography{eetc}

		\end{document}